\documentclass[fleqn,usenatbib]{mnras}
\usepackage{newtxtext,newtxmath}
\usepackage[T1]{fontenc}
\usepackage{ae,aecompl}

\usepackage{graphicx}	
\usepackage{amsmath}	

\usepackage{url}            
\usepackage{booktabs}       
\usepackage{amsfonts}       
\usepackage{nicefrac}       
\usepackage{microtype}      
\usepackage{lipsum}
\usepackage{hyperref}       
\usepackage{cleveref}       

\usepackage{verbatim}

\usepackage{tikz}
\usepackage{mathdots}
\usepackage{cancel}
\usepackage{siunitx}
\usepackage{array}
\usepackage{multirow}
\usepackage{gensymb}
\usepackage{tabularx}
\usepackage{booktabs}
\usepackage[toc,page]{appendix}
\usepackage{xcolor}
\usepackage{times}

\newcommand{\correction}[1]{\textcolor{black}{#1}}
\newcommand{\mtrv}[1]{{\color{black}{#1}}}
\newcommand{\mtrvv}[1]{{\color{black}{#1}}}
\newcommand{\Baojiu}[1]{\textcolor{black}{#1}}

\newcommand{\Carlton}[1]{\textcolor{black}{#1}}

\newcommand{\Takahiro}[1]{\textcolor{black}{#1}}
\newcommand{\Mpch}{\,h^{-1}{\rm Mpc}}

\title[non-Gaussian RSD model]{Towards a non-Gaussian model of redshift space distortions}
\author[C. Cuesta-Lazaro et al.]{Carolina Cuesta-Lazaro\thanks{E-mail: carolina.cuesta-lazaro@durham.ac.uk}$^{1,2}$,
Baojiu Li$^1$,
Alexander Eggemeier$^1$,
Pauline Zarrouk$^1$,
\newauthor
Carlton M. Baugh$^{1,2}$,Takahiro Nishimichi$^{3,4}$,
and Masahiro Takada$^{4}$
\\
$^{1}$Institute for Computational Cosmology, Department of Physics, Durham University, South Road, Durham DH1 3LE, UK\\
$^{2}$Institute for Data Science, Durham University, South Road, Durham DH1 3LE, UK\\
$^{3}$Center for Gravitational Physics, Yukawa Institute for Theoretical Physics, Kyoto University, Kyoto 606-8502, Japan\\
$^{4}$Kavli Institute for the Physics and Mathematics of the Universe (Kavli IPMU, WPI), \\
The University of Tokyo Institutes for Advanced Study, The University of Tokyo, Kashiwa, Chiba 277-8583, Japan
}

\date{Accepted XXX. Received YYY; in original form ZZZ}

\pubyear{2019}

\begin{document}

\label{firstpage}
\pagerange{\pageref{firstpage}--\pageref{lastpage}}
\maketitle


\begin{abstract}
To understand the nature of the accelerated expansion of the Universe, we need to combine constraints on the expansion rate and growth of structure. The growth rate is usually extracted from three dimensional galaxy maps by exploiting the effects of peculiar motions on galaxy clustering. However, theoretical models of the probability distribution function (PDF) of galaxy pairwise peculiar velocities are not accurate enough on small scales to reduce the error on
\Baojiu{theoretical predictions} to the level required to match the precision expected for measurements from future surveys. \Baojiu{Here, we} improve the modelling of the pairwise velocity distribution by using the Skew-T PDF, which has nonzero skewness and kurtosis. 
Our model \Baojiu{accurately} reproduces the redshift-space multipoles (monopole, quadrupole and hexadecapole) predicted by N-body simulations, above scales of about $10\Mpch$. We illustrate how a Taylor expansion of the streaming model can reveal the contributions of the different moments to the clustering multipoles, which are independent of the shape of the velocity PDF. 
\Baojiu{The Taylor expansion explains why the Gaussian streaming model works well in predicting the first two redshift-space multipoles, although the velocity PDF is non-Gaussian even on large scales. Indeed,} any PDF with the correct first two moments would produce precise results for the monopole down to scales of about $10\Mpch$, and for the quadrupole down to about $30\Mpch$. An accurate model for the hexadecapole needs to include higher-order moments.
\end{abstract}






\section{Introduction}

By combining different experiments, such as the supernovae standard candles \citep{1997ApJ...483..565P, 1998AJ....116.1009R,2007ApJ...659...98R}, and the cosmic microwave background (CMB) temperature anisotropies \citep{2018arXiv180706209P}, astronomers have inferred that the expansion of the Universe is accelerating. If general relativity is the correct theory of gravity, concordance with the aforementioned experiments requires that about $95\%$ of the energy content of the universe is invisible to us, since it does not exhibit electromagnetic interactions. Most of this energy is postulated to be in the form of a cosmological constant, $\Lambda$, which drives the observed accelerated expansion of the Universe. Its nature is, however, unknown. 

\Baojiu{A} compelling theoretical explanation attributes the cosmological constant to the energy of empty space. Nonetheless, its estimated value \citep{Carroll2001} has produced the largest discrepancy between theoretical predictions and observations ever encountered in physics. \Baojiu{Therefore, d}espite the success of the standard model of cosmology (known as $\Lambda$CDM) in explaining observations, we \Baojiu{still} do not understand the nature of \Baojiu{some of} its constituents, which has led cosmologists to look for evidence beyond $\Lambda$CDM.

\Baojiu{Most of the} extensions to the standard model add new degrees of freedom to either the energy content of the Universe, or to the way space-time geometry reacts to it, by modifying general relativity \citep{2012PhR...513....1C, 2016ARNPS..66...95J}. The most popular example of extra degrees of freedom in the energy content is dynamical dark energy, that allows for temporal variation of the dark energy equation of state, driven by a scalar field. The second class of models involve modifications to the Einstein-Hilbert action, and are commonly denoted as modified gravity. The modified action adds new terms to the Poisson equation, which are  identified with the appearance of an additional fifth force. However, the boundary between the two categories is not \Baojiu{always} well defined. 
 
To distinguish between these different scenarios observationally, tests of the background expansion \Baojiu{history} are not sufficient. Viable gravity theories can be tuned to reproduce the observed evolution of the scale factor with time, and therefore are indistinguishable on the background level. However, by including information about the rate at which cosmic structures grow, we can detect modifications to gravity. The growth of cosmic structure is the outcome of a competition between the expansion of the Universe and the gravitational pull, generated by inhomogeneities. If there is an additional fifth force, but the expansion is compatible with that observed, the rate at which structures in the Universe grow will be modified. 

To estimate the large-scale growth of structure in the Universe, we look at the statistics of 3-D galaxy maps. These maps contain the angular position of galaxies on the sky, together with their redshifts. Assuming that galaxies are at rest, as the photons they emit travel towards us through an expanding Universe, their wavelengths stretch accordingly. Therefore, we observe the redshifted light of distant galaxies. We can translate this redshift into a comoving distance by introducing the Hubble factor, $H(z)$,
\begin{equation}
    r(z) = \int_0^z \frac{d z'}{H(z')}
    \label{eq:distance2redshift}
\end{equation}
where $r(z)$ is the comoving distance to the galaxy\Baojiu{, and we have used the natural unit where the speed of light $c=1$}.

However, galaxies also move due to the gravitational pull generated by the inhomogeneous distribution of matter around them. If a source that emits light moves, the wavelength of the light gets further redshifted due to the Doppler effect. If we ignored this effect then we would infer the wrong distance, $\mathbf{s}$, given by,
\begin{equation}
\mathbf{s} = \mathbf{r} + \frac{\mathbf{v}(\mathbf{r})\hat{z}}{\mathcal{H}}\hat{z},
\label{eq:rspositions}
\end{equation}
instead of the real position of the galaxy, $\mathbf{r}$ ,where $\mathbf{v}(\mathbf{r})$ is the peculiar velocity of the galaxy, $\mathcal{H} = a H(a)$ the comoving Hubble factor, and the inferred distance, $\mathbf{s}$, is the redshift space distance. Note that we have assumed the observer is far away from the sources, and therefore the line-of-sight direction can be fixed to a particular direction, which we arbitrarily set as the $\hat{z}$ axis. This approximation, known as the plane-parallel approximation, has been so far given results that lie within the statistical error bars of current surveys \citep{2012MNRAS.420.2102S}. However, this approximation will need to be used more carefully in the analysis of upcoming surveys.
 
The translation between redshift and distance is in reality then more complex than Eq.~(\ref{eq:distance2redshift}), since we need to disentangle the combination of the galaxy's position and its peculiar velocity along the line of sight.  However, this complication ends up being beneficial because peculiar velocities are generated by the gravitational pull of the inhomogeneous matter distribution, and therefore allow us \Baojiu{to extract information on the latter. In particular, as mentioned above, it is possible to detect the existence or constrain the strength of fifth forces by studying the growth of structure inferred from the statistics of the peculiar velocity field}, \correction{see e.g. \citep{Guzzo:2008ac, 2017JCAP...08..029B, 2019MNRAS.485.2194H}}.

To extract the growth rate, we measure the effect of peculiar velocities on clustering known as Redshift Space Distortions (RSD). The amount of clustering is quantified as the probability of finding a pair of galaxies at a given \Baojiu{separation}, compared to the probability for a random distribution of galaxies. This statistic is known as the 
two-point correlation function, defined as,
\begin{equation}
\xi^R (r) = \langle \delta(\mathbf{x}) \delta(\mathbf{x} + \mathbf{r}) \rangle,
\end{equation}which, due to statistical isotropy and homogeneity, only depends on pair separation, $r$. The two-point correlation function is a useful statistic because it summarises all the statistical information contained in a Gaussian random field. Even though the evolved density field is non-Gaussian, it can still be used to constrain the cosmological parameters. However, the higher-order correlation functions of the evolved density field are non-zero, and hence contain additional independent information over and above that encoded within the variance \citep[see e.g., ][in the context of redshift-space clustering]{Song15,Gagrani17}. 

Due to the peculiar motions of galaxies, we observe redshift space positions, $\mathbf{s}$, instead of the real space positions, $\mathbf{r}$, and thus we can only measure, 
\begin{equation}
\xi^S (s_\perp, s_\parallel) = \langle \delta(\mathbf{x}) \delta(\mathbf{x} + \mathbf{s}) \rangle,
\label{eq:redshift_tpcf}
\end{equation}
which depends on both the pair separation, $s$, and its inclination with respect to the line-of-sight direction. Throughout we denote separations perpendicular and parallel to the line of sight as $s_\perp$ and $s_\parallel$, respectively.

The so-called redshift space correlation function, $\xi^S (s_\perp, s_\parallel) $, is a combination of both real space clustering, $\xi^R(r)$,  and the probability of finding a pair of galaxies with a given relative velocity along the line of sight, also denoted as pairwise velocity distribution,  as we will show in Section~\ref{sec:intro_streaming} using the Streaming Model of RSD, \correction{ see e.g. \citep{1995ApJ...448..494F, 2004PhRvD..70h3007S}}. Since clustering in redshift space is affected by relative peculiar motions, it contains information about the growth of structure.

On this paper, we focus on improving the accuracy of models of the redshift space correlation function, to improve our estimates of the growth rate. The main hurdle that has to be overcome, is the non-linear evolution of the density and velocity fields produced by non-linearities in the continuity and Euler equations that drive gravitational collapse. As we will see, this is particularly relevant to describe the mapping of pairs from real to redshift space, which is necessary to model the two-point correlation function in Eq.~\ref{eq:redshift_tpcf}. The development of this mapping is our focus here. 

State-of-the-art constraints on the growth factor are found measuring the two-point correlation function in redshift space (e.g., \citealt{2017MNRAS.469.1369S} for galaxies from BOSS, \citealt{10.1093/mnras/sty506} for eBOSS quasars), which have reported growth factors consistent with general relativity. The authors in  \cite{2017MNRAS.469.1369S}, used measurements of the two-point correlation function down to separations of $25\Mpch$, beyond which theoretical predictions introduce larger systematic errors than the statistical errors of the measurement itself, thus biasing the estimate of the growth factor.  For future surveys, the expected statistical errors will be significantly smaller (e.g., \citealt{Huterer:2013xky}), and so we will need more accurate theoretical predictions down to small scales than those used in the analysis of current surveys, to improve constraints on the growth rate, and to avoid making catastrophic errors of interpretation \citep{Jennings:2011}. 

The goal of this paper is threefold. Firstly, we introduce an extension to the simplest Streaming Model, that assumes Gaussian relative motions, that improves the accuracy of theoretical predictions for the clustering multipoles. Secondly, we show a comparison of state-of-the-art models for the streaming model ingredients with high resolution N-body simulations. Finally, we analyse the effect of the different velocity moments on the clustering multipoles, and assess how accurate their theoretical predictions need to be for an RSD model that is at least as accurate as the measurements from future surveys.

This paper is structured as follows. In Section~2, we summarise the theoretical framework used to model the RSD effect on the two-point correlation function, the streaming model, and we present the Skew-T (hereafter ST) distribution as a model for the pairwise velocity distribution that includes non-Gaussian features. In Section~3 we compare the accuracy of the ST model with a simple Gaussian distribution, using N-body simulations. In Section~4, we study a Taylor expansion of the streaming model integrand, which allows us to examine how the different higher-order moments affect the multipoles. In Section~5 we analyse the sensitivity of the multipoles to the different real space ingredients of the streaming model. Finally, in Section~6 we summarise our main results and draw our conclusions.

\section{The streaming model of redshift space distortions}
\label{sec:intro_streaming}
The streaming model provides the mapping from the real-space two-point correlation function to the observed anisotropic \Baojiu{two-point} correlation function in redshift space. Since objects viewed in redshift space are the same as those in real space, but have been moved to different positions, we can relate their density contrasts by imposing mass conservation
\begin{equation}
\left(1 + \delta^S(\mathbf{s})\right) \mathrm{d}^3\mathbf{s} = \left(1 + \delta^R(\mathbf{r})\right)\mathrm{d}^3\mathbf{r},
\end{equation}
where superscript $S$ denotes redshift space, and $R$ real space. This expression can be further manipulated 
 \citep{2004PhRvD..70h3007S} to obtain a relation between real and redshift space clustering,
\begin{equation}
    1 + \xi^S\!(s_\perp, s_\parallel) = \int_{-\infty}^{\infty}\! \mathrm{d}r_\parallel~ \left(1 + \xi^R (r)\right) \mathcal{P}(v_\parallel = s_\parallel - r_\parallel | \mathbf{r}),
    \label{eq:streaming}
\end{equation}
where $r^2 = r_\parallel ^2 + r_\perp ^2$,  $s_\perp = r_\perp$\Baojiu{,  $\mathcal{P}(v_\parallel|{\bf r})$ is the pairwise velocity distribution, and $v_\parallel = v_{\parallel,1} - v_{\parallel,2}$, is the line of sight relative velocity of the pair of tracers. In our} convention, $v_\parallel$ is defined as negative (positive) if the pairs \Baojiu{are} approaching (receding \Baojiu{from}) \Baojiu{each other}. Eq.~\eqref{eq:streaming} \Baojiu{is} known as the streaming model \citep{1995ApJ...448..494F}, which
simply
\Baojiu{states} that the probability of finding a pair of objects at a distance $\mathbf{s}$ in redshift space is given by the sum over all possible real space distances, $\mathbf{r}$, that would make us infer the redshift space position.  While the streaming model is one way to move forward, fully Eulerian perturbation theory treatments based on the same expression can also accurately describe redshift space clustering \citep{2010PhRvD..82f3522T}.

The plus one terms in Eq.~(\ref{eq:streaming}) ensure that given a universe with randomly placed galaxies, if the pairwise velocity distribution is dependent on the pair separation, then we would still observe clustering in redshift space induced by the coherent velocity field. If, however, the pairwise velocity distribution does not depend on pair separation, the plus one terms on both sides in Eq.~(\ref{eq:streaming}) cancel out.  

Note that Eq.~(\ref{eq:streaming}) is exact, \Baojiu{and} the only approximation we have made so far is the plane-parallel approximation to select a particular line of sight. Nonetheless, the apparent simplicity of the streaming model may be deceptive, \Baojiu{as} the complexity of the gravitational dynamics is hidden in the shape of $\mathcal{P}(v_\parallel|{\bf r})$ and its dependence on pair separation. Broadly speaking, on small scales within dark matter haloes, virial motions produce a large velocity dispersion that reduces the amount of clustering along the line of sight; the size of this effect increases with halo mass. On larger scales, galaxies in-falling into larger structures shift the mean velocity to negative values, producing a change in the opposite sense to those on small scales, that increases the inferred clustering along the line of sight \citep{Kaiser87}. 

It has been known for a long time \citep{2004PhRvD..70h3007S} that this scenario is further complicated by the non-Gaussian nature of the pairwise velocity distribution, which is evident from its non-zero skewness and kurtosis. There is no Gaussian limit for pairwise velocities on large scales, since velocity differences cancel out long-range contributions and leave only the local, nonlinear component of the velocity at the two different locations. Here, we focus on extending the streaming model to include these non-Gaussian features, as predicted by N-body simulations.

\begin{figure}
\centering
\tikzset{every picture/.style={line width=0.75pt}} 

\begin{tikzpicture}[x=0.75pt,y=0.75pt,yscale=-1,xscale=1]

\draw (237.96,152.94) node  {\includegraphics[width=44.65pt,height=42.42pt]{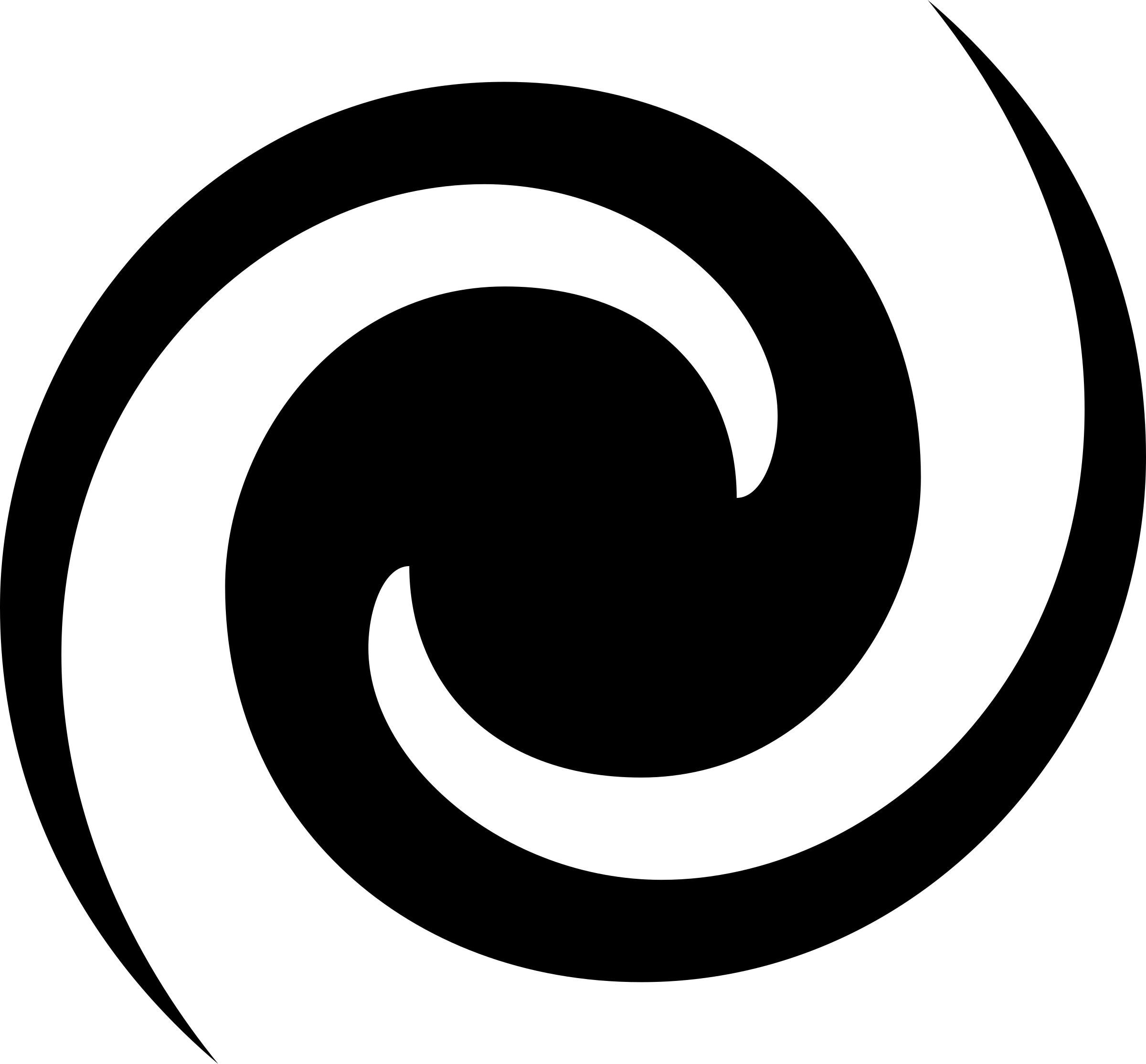}};
\draw  [draw opacity=0][fill={rgb, 255:red, 126; green, 211; blue, 33 }  ,fill opacity=0.44 ] (281.92,136.36) .. controls (283.23,141.07) and (283.28,146.02) .. (282.22,150.91) -- (241.54,150.31) -- cycle ; \draw   (281.92,136.36) .. controls (283.23,141.07) and (283.28,146.02) .. (282.22,150.91) ;
\draw (385.52,102.45) node [rotate=-34.85] {\includegraphics[width=44.65pt,height=42.42pt]{figures/galaxy.png}};
\draw    (385.07,103.31) -- (446.49,80.98) ;
\draw [shift={(448.37,80.3)}, rotate = 520.02] [color={rgb, 255:red, 0; green, 0; blue, 0 }  ][line width=0.75]    (10.93,-3.29) .. controls (6.95,-1.4) and (3.31,-0.3) .. (0,0) .. controls (3.31,0.3) and (6.95,1.4) .. (10.93,3.29)   ;

\draw    (385.07,103.31) -- (405.43,151.28) ;
\draw [shift={(406.21,153.12)}, rotate = 247] [color={rgb, 255:red, 0; green, 0; blue, 0 }  ][line width=0.75]    (10.93,-3.29) .. controls (6.95,-1.4) and (3.31,-0.3) .. (0,0) .. controls (3.31,0.3) and (6.95,1.4) .. (10.93,3.29)   ;

\draw [line width=2.25]    (241.54,150.31) -- (240.27,96.99) ;
\draw [shift={(240.17,92.99)}, rotate = 448.64] [color={rgb, 255:red, 0; green, 0; blue, 0 }  ][line width=2.25]    (17.49,-5.26) .. controls (11.12,-2.23) and (5.29,-0.48) .. (0,0) .. controls (5.29,0.48) and (11.12,2.23) .. (17.49,5.26)   ;

\draw  [color={rgb, 255:red, 155; green, 155; blue, 155 }  ,draw opacity=1 ] (379.89,103.31) .. controls (379.89,100.83) and (382.21,98.81) .. (385.07,98.81) .. controls (387.93,98.81) and (390.24,100.83) .. (390.24,103.31) .. controls (390.24,105.79) and (387.93,107.8) .. (385.07,107.8) .. controls (382.21,107.8) and (379.89,105.79) .. (379.89,103.31) -- cycle ; \draw  [color={rgb, 255:red, 155; green, 155; blue, 155 }  ,draw opacity=1 ] (381.41,100.13) -- (388.73,106.49) ; \draw  [color={rgb, 255:red, 155; green, 155; blue, 155 }  ,draw opacity=1 ] (388.73,100.13) -- (381.41,106.49) ;
\draw  [dash pattern={on 0.84pt off 2.51pt}]  (235.02,153.12) -- (385.07,103.31) ;

\draw  [dash pattern={on 0.84pt off 2.51pt}]  (241.54,150.31) -- (305.03,149.71) ;

\draw (451.06,95.88) node    {$\hat{r}$};
\draw (384.62,139.32) node    {$\hat{t}$};
\draw (385.9,69.05) node  [color={rgb, 255:red, 128; green, 128; blue, 128 }  ,opacity=1 ]  {$\hat{n}$};
\draw (245.1,71.82) node    {$los\ =\ \hat{z}$};
\draw (339.19,205.03) node    {$\hat{t} \ =\ \hat{r} \times \hat{n}$};
\draw (337.29,177.71) node    {$\hat{n} \ =\ \hat{r} \times \hat{z}$};
\draw (281.72,119.66) node    {$\frac{\pi }{2} \ -\theta $};

\end{tikzpicture}

\caption{Decomposition of the three dimensional distance vector into a radial component along the pair distance, $\mathbf{\hat r}$,  a normal component, $\mathbf{\hat n}$,  which is perpendicular to both the line of sight direction and the pair separation vector, and a transverse component, $\mathbf{\hat t}$,  which completes the basis formed by the radial and normal vectors. After projecting the distance vector onto the line of sight, only the radial and transverse component will give a non-zero contribution.}
\label{fig:diagram}
\end{figure}

Throughout, we will use the relation between the full three dimensional pairwise velocity and its line-of-sight projection. The line-of-sight pairwise velocity distribution can be obtained by integrating the full distribution $\mathcal{P}(v_r, v_t | r)$, where the radial velocity, $v_r$, and the transverse velocity, $v_t$, are defined as the velocity \Baojiu{components} parallel and transverse to the pair 
\Baojiu{separation vector}, respectively. Due to \Baojiu{statistical} isotropy, we 
\Baojiu{only need to} select one component 
from
the two dimensional transverse velocity. For ease of computation, we chose the one which will contribute to the line-of-sight projection, \Baojiu{i.e., the one in the plane spanned by the galaxy pair and the observer;} see Fig.~\ref{fig:diagram}. Thus,
\begin{equation}
    v_\parallel = v_r \cos\theta + v_t \sin\theta , 
\end{equation}
where $\theta$ is the angle between the pair 
\Baojiu{separation} vector and the line of sight, $\theta = \tan ^{-1} \left( {r_\perp}/{r_\parallel} \right)$. Therefore, 
\begin{equation}
    \mathcal{P}(v_\parallel | r_\perp, r_\parallel) = \int\! \frac{\mathrm{d} v_r}{\sin\theta}~ \mathcal{P}\!\left(v_r, v_t = \frac{v_\parallel - v_r \cos \theta}{\sin \theta}\bigg|r \right) .
    \label{eq:projection}
\end{equation}
The relations between the moments of the two distributions are given by,
\begin{equation}
c_n (r_\perp, r_\parallel) = \sum_{k=0}^n {n\choose k} \mu^k (1 - \mu^2)^{\frac{n -k}{2}} c_{k, n-k}(r),
\end{equation}
where $c_n$ denotes the $n$-th central moment of the line of sight projected distribution, $ \mathcal{P}(v_\parallel | r_\perp, r_\parallel)$,  and $c_{k, n-k}$ the $k$-th radial moment, $(n-k)$-th transverse moment of $\mathcal{P}(v_r, v_t |r ) $, \correction{and $\mu = \cos \theta$}. The $n$-th moment about the origin is denoted as $m_n$.

\subsection{The Gaussian Streaming Model}

The \Baojiu{commonly used} model for the redshift space correlation function is known as the Gaussian streaming model \citep[GSM;][]{1995ApJ...448..494F,2011MNRAS.417.1913R}. The radial and transverse components of the pairwise velocity are taken to be independently Gaussian distributed. The line-of-sight projection can therefore be written as
\begin{equation}
    \mathcal{P}_\mathrm{G}(v_\parallel | \mathbf{r}) =
     \frac{1}{\sqrt{2 \pi \sigma_{12}^2(\mathbf{r})}} \exp\left[{-\frac{\left( v_\parallel - v_{12}(\mathbf{r}) \right)^2}{2  \sigma_{12}^2(\mathbf{r})}}\right],
\end{equation}
where $v_{12}(\mathbf{r})$, denoted as $m_1(\mathbf{r})$ in our notation,  and $\sigma_{12}(\mathbf{r})$, equivalent to $\sqrt{c_2(\mathbf{r})}$, are the projections of the radial and transverse moments onto the line of sight, and are both dependent on the pair separation.

As explained in the previous section, a Gaussian distribution does not accurately describe the pairwise velocity distribution for an evolved matter distribution even \Baojiu{for} large pair separations. However, this simplified assumption gives an accurate description of the clustering of dark matter haloes on scales larger than $30\Mpch$ \citep{2011MNRAS.417.1913R, 2014MNRAS.437..588W}. Later on, we shall illustrate how the accuracy of this model stems from the integral in Eq.~\eqref{eq:streaming} over the pairwise velocity distribution, which on large scales only receives contributions from the lowest order pairwise velocity moments.

Nonetheless, an accurate model on smaller scales requires non-vanishing higher order moments, mainly the skewness and kurtosis. Different approaches have been taken towards such a model in the literature. On one hand \cite{2015PhRvD..92f3004U} performed an Edgeworth expansion around a Gaussian distribution to add skewness, and found improvements with respect to the Gaussian streaming model on scales smaller than $30\Mpch$. \correction{We provide a more in-depth discussion of this model on the following section.}
On the other hand, \Baojiu{a number of authors} \citep[e.g.,][]{1996MNRAS.279.1310S, 2007MNRAS.374..477T, 2015MNRAS.446...75B,2016MNRAS.463.3783B, 2018MNRAS.479.2256K} have all used mixtures of normal or quasi-normal distributions to model a skewed and heavy-tailed distribution. The first approach by \cite{1996MNRAS.279.1310S} modelled the motion of the halo centre of mass using a Maxwellian distribution which is then weighted by the Press-Schechter mass function. \cite{2007MNRAS.374..477T} developed a similar approach using the halo model \citep{2002PhR...372....1C}, but assuming that, at fixed environmental density around the halo pair, the pairwise velocity distribution of halos is Gaussian. The skewness is then developed by weighting these Gaussian distributions with the probability of finding a given density. The parameters of the model are calibrated \Baojiu{using} N-body simulations.

Further developments were introduced by \cite{2015MNRAS.446...75B}, who replaced the mixing distribution described above by another Gaussian, which assumes that the mean and standard deviations of the "local" Gaussian distributions are themselves jointly distributed according to a bivariate Gaussian. This model, however, cannot generate distributions that are sufficiently skewed to explain the halo pairwise velocity distribution. This limitation was later overcome by performing an Edgeworth expansion on the local distributions, which added skewness to the Gaussian distribution \citep{2016MNRAS.463.3783B}. 

\Baojiu{A more recent study} 
by \citet{2018MNRAS.479.2256K} used a generalised hyperbolic distribution (GHD) to model the pairwise velocity distribution of N-body simulations. 
\Carlton{In this case} 
the relation between the parameters of the distribution and velocity moments as a function of pair 
\Baojiu{separation} is not given, \Baojiu{and} the model 
\Carlton{requires}
five free parameters with a 
two dimensional dependency on the pair 
\Baojiu{separation} vector. 
\subsection{The Edgeworth Streaming Model}

\correction{The Edgeworth Streaming Model introduced by \cite{2015PhRvD..92f3004U}, is one of the simplest extensions to the Gaussian Streaming Model. The authors used an Edgeworth expansion of the velocity PDF to extend the validity of the Gaussian Streaming Model towards smaller scales. The Edgeworth expansion is an asymptotic series expansion to a probability density function, which implies that there is no guarantee of convergence when more terms are added to the expansion. See \cite{Sellentin:2017aii} for an interesting discussion on the Edgeworth expansion and its applications to cosmology.}

\correction{Expanding the line of sight velocity PDF around a Gaussian distribution one obtains, to first order,}

\begin{equation}
  \begin{split}
    \mathcal{P}_\mathrm{E}(v_\parallel | \mathbf{r})\; &=
     \frac{1}{\sqrt{2 \pi \sigma_{12}^2(\mathbf{r})}} \exp\left[{-\frac{\left( v_\parallel - v_{12}(\mathbf{r}) \right)^2}{2  \sigma_{12}^2(\mathbf{r})}}\right] \\ &\times\,\left(1 + \frac{\Lambda_{12}}{6 \sigma_{12}^3} H_3\left(\frac{ v_\parallel - v_{12}}{\sigma_{12}}\right)\right),
 \end{split}
\end{equation}

\correction{where $\Lambda_{12}$ is the third order cumulant of the velocity PDF projected onto the line of sight direction, and $H_3$ the third order probabilists'  Hermite polynomial, $H_3 (x)= x^3 - 3x$.}

\correction{In the next section, we present a flexible model that we believe is simpler than the ones mentioned above and achieves similar or better levels of accuracy.}

\subsection{The Skewed Student-t (ST) Streaming Model}

A study of the cluster-galaxy cross correlation 
by \citet{10.1093/mnras/stt411} found that the skewed Student-t distribution \citep[ST;][]{2009arXiv0911.2342A} gives an accurate description of the \Baojiu{cluster-galaxy} pairwise velocity statistics \Baojiu{predicted by} simulations. 
\Carlton{The main advantage of using this distribution to model RSD}
is that its parameters can be written as functions of the 
four \Baojiu{lowest-}order moments. 
\Carlton{Here}
we use the ST distribution to model the redshift-space clustering \Baojiu{of galaxy or halo pairs} on all scales. 

In recent years, there has been increasing interest in 
such flexible probability density functions that can accommodate different degrees of skewness and kurtosis. More specifically, 
a successful approach \Baojiu{proposed} by \cite{2009arXiv0911.2342A}, 
found that a skewed, multi-variate, distribution can be generated by combining a symmetric density function with a cumulative distribution function as follows
\begin{equation}
	\label{eq:symmetry}
    f(x) = 2 \, f_0(x) \, G\left(w(x)\right), \; x \in \mathbb{R}^d,
\end{equation}
where $f_0(x)$ is a symmetric PDF defined in $\mathbb{R}^d$, $G$ is a one-dimensional cumulative distribution function, whose derivative satisfies $G^{'} (x) = G^{'} (-x)$, and $w$ is a real-valued odd function in $\mathbb{R}$.

Since we are interested in a distribution that 
\Carlton{displays} 
both skewness and extended tails, the symmetric function $f_0$ can be chosen to be a Student's $t$-distribution, hereafter 
\Carlton{referred to as the $t$-distribution,}
which in one dimension is given by
\begin{equation}
    \label{eq:student}
    f_0(x) = t_1(x - x_c |w, \nu) := \frac{\Gamma\left(\frac{\nu + 1}{2} \right)}{\sqrt{\nu \pi} w \Gamma(\frac{\nu}{2})} 
    \left( 1 + \frac{1}{\nu} \left( \frac{x - x_c}{w}\right)^2 \right)^{-\frac{\nu+1}{2}}.
\end{equation}
The $t$-distribution is characterised by three parameters: the location $x_c$, the shape parameter, $w$, and the number of degrees of freedom, $\nu$. The latter controls the 
\Carlton{decay of probability}
in the tails, and 
\Carlton{therefore allows us} 
to describe distributions with varying degrees of kurtosis.

The skewed multi-variate distribution which originates from the $t$-distribution by using Eq.~\eqref{eq:symmetry} is known as the skew-$t$ distribution, hereafter ST. Its density function for a one dimensional random variable, $x$, is,
\begin{equation}
\begin{split}
     f_{\text{ST}}&(x| x_c, w, \alpha, \nu)  := \\ & \frac{2}{w} t_1(x-x_c | 1, \nu)T_1\left[\alpha \frac{(x - x_c)}{w} \left( \frac{\nu + 1}{\nu + \left(\frac{x-x_c}{w}\right)^2}\right)^{1/2} ; \nu + 1 \right],
    \label{eq:azzalini}
\end{split}
\end{equation}
where $t_1$ is the one dimensional $t$-distribution defined by Eq.~\eqref{eq:student}, and $T_1$ is the one dimensional cumulative $t$-distribution with $\nu + 1$ degrees of freedom. The ST distribution has an extra skewness parameter, $\alpha$, compared to the $t$-distribution.

Note that the dependence of the distribution parameters on the pair separation vector, $\mathbf{r}$, has been omitted for clarity. The relation between these parameters and the \Carlton{four lowest} order moments of the ST distribution can be found in \Baojiu{Appendix} \ref{app:moments_st}.

\section{Comparison with N-body simulations}

In this section, we assess the performance of the different RSD models by comparing them to a set of dark matter only $\Lambda$CDM simulations.

\subsection{\mtrv{Simulations}}
We use the Dark Quest \citep{2018arXiv181109504N} set of simulations, which consists of fifteen independent realisations \Carlton{of the density fluctuations in a cosmological volume, adopting}
the 
\Carlton{best-fitting}
cosmological parameters given by \Carlton{the} Planck CMB data \citep{2016A&A...594A..13P}  
\begin{equation}
\begin{split}
  \{\omega_b, \omega_c, \Omega_{\text{DE}}&,  \ln(10^{10} A_s), n_s,  w_{\mathrm{DE}} \} =    \\ &\{0.02225,0.1198,0.6844,3.094,0.9645,-1\},
    \label{eq:parameters}
\end{split}
\end{equation}
where $\omega_b\equiv \Omega_b h^2$, and  $\omega_c\equiv \Omega_c h^2$ are the physical density parameters of baryons and cold dark matter, \Carlton{respectively,} $\Omega_{\text{DE}}=1-(\omega_b + \omega_c + \omega_\nu)/h^2$ is the dark energy density parameter (assuming a flat Universe \Takahiro{and the neutrino density parameter, $\omega_\nu$ corresponding to the total mass of $0.06$eV for the three neutrino species}), $A_s$ and $n_s$ are the amplitude and tilt of the primordial curvature power spectrum normalised at $0.05 \,\, \mathrm{Mpc}^{-1}$, and $\omega_{\mathrm{DE}}$ is the equation of state parameter of dark the energy. 

The simulations follow the evolution of $2048^3$ particles in a \Takahiro{comoving} box of size $L=2h^{-1}$Gpc, which translates into a particle mass of $m_p = 8.158 \times 10^{10} h^{-1} M_\odot$, using the Tree-Particle Mesh code \Baojiu{{\sc gadget}2} \citep{10.1111/j.1365-2966.2005.09655.x}. Halo catalogues were constructed using the publicly available \Baojiu{{\sc rockstar}} halo finder \citep{2013ApJ...762..109B}. 
\Carlton{Here,} we focus on accurate predictions for massive central halos, with masses above $10^{13}\Baojiu{h^{-1}}M_\odot$, and 
leave the predictions for galaxies 
\Carlton{to} 
future work. In all figures below, we show the mean simulation measurements over the fifteen independent realisations \Carlton{of the cosmological volume}, \correction{with errorbars representing one standard deviation of the mean measurements}. 
All results are shown for the $z=0$ snapshots.

\subsection{The 
ingredients of the streaming model}

The streaming model, Eq.~\eqref{eq:streaming}, takes as input both the pairwise velocity distribution and the real space two-point correlation function. In this subsection we will show the measurements of both 
\Baojiu{ingredients from the simulations}, together with \Baojiu{their} theoretical predictions 
\Baojiu{for the given} cosmological parameters.

\begin{figure*}
    \centering
    \includegraphics[width=0.9\textwidth]{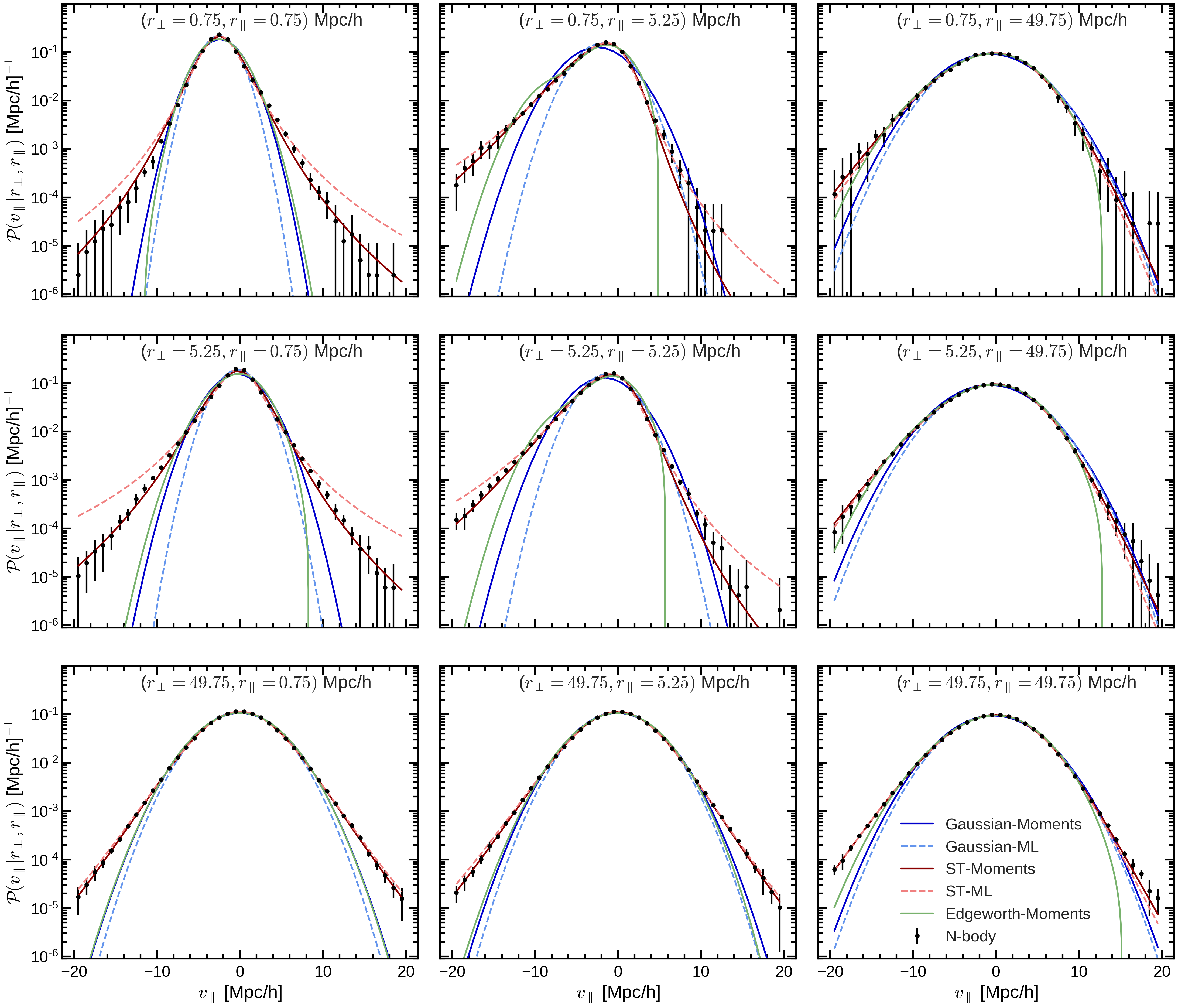}
    \caption{The pairwise line of sight velocity distribution for massive dark matter halos in the simulation at $z = 0$, evaluated at different pair-separations. Columns show increasing $r_\parallel$ separation, whilst rows show increasing $r_\perp$. The black dots show the mean measurements from the N-body simulation and their standard deviation, whilst the solid (dashed) curves show the different models found using the method of moments (maximum likelihood) estimate. The Gaussian model is shown in blue, \correction{ the Edgeworth expansion model is shown in green,} and the ST model, that includes skewness and kurtosis, is shown in red.}
    \label{fig:line_of_sight_pdf}
\end{figure*}

Using the halo catalogues \Baojiu{from the simulations}, we measure the pairwise velocity distribution in bins of $0.5\Mpch$ size \Baojiu{(note that the velocities are rescaled by $\mathcal{H}$ so that they have the unit of length). As mentioned above, in our convention} the pairwise velocity is defined as negative (positive) when the members of the pair are approaching (receding \Baojiu{from) each other}. We show the measured pairwise distribution from the simulations, \Baojiu{for a few selected cases of $(r_\perp,r_\parallel)$,} in Fig.~\ref{fig:line_of_sight_pdf}. The figure shows increasing $r_\parallel$ values from left to right, and increasing $r_\perp$ from top to bottom.

\Baojiu{The black dots in Fig.~\ref{fig:line_of_sight_pdf}} 
\Carlton{show} 
the measured pairwise velocity of dark matter haloes, while the lines give the ST (red) and Gaussian (blue) distributions obtained by applying two different methods to find the 
\Carlton{best-fit} parameters, which will be described in Section~\ref{sec:parameters}. It is evident that in all cases the Gaussian distributions are a poorer fit to the simulation measurements 
\Carlton{than}
the ST distributions. In particular, by comparing the symbols with the blue curves, we note that for all pair separations there is a significant kurtosis in the simulation data which a Gaussian distribution fails to capture. 

In the cases of $r_\parallel=5.25\Mpch$ and $r_\perp=0.75$ or $5.25\Mpch$, the pairwise velocity distributions are also very strongly skewed towards negative $v_\parallel$, which is because such close halo pairs are more likely to be found in high-density regions where haloes approach each other ($v_\parallel<0$), than in void regions where haloes tend to move apart ($v_\parallel>0$) . The skewness, however, decreases for much larger $r_\perp$ (e.g. $49.75\Mpch$, the bottom panel of the central column of Fig.~\ref{fig:line_of_sight_pdf}) or $r_\parallel$ (the right column of Fig.~\ref{fig:line_of_sight_pdf}), because the probabilities of infalling and receding halo pairs tend to \mtrv{be} even out for large separations. On the other hand, the left column \Carlton{of Fig.~\ref{fig:line_of_sight_pdf}} shows that for very small $r_\parallel$ (e.g., $0.75\Mpch$), the skewness is small again, which is because in this case the pair separation vector is nearly perpendicular to the line of sight, and $v_\parallel\approx v_t$,  meaning that $v_\parallel$ has equal probability to be in 
\Carlton{any}
direction within the plane perpendicular to the pair separation vector due to statistical isotropy, that is, equal probability of $v_\parallel>0$ and $v_\parallel<0$.

\begin{figure*}
    \centering
    {
    \includegraphics[width=0.35\textwidth]{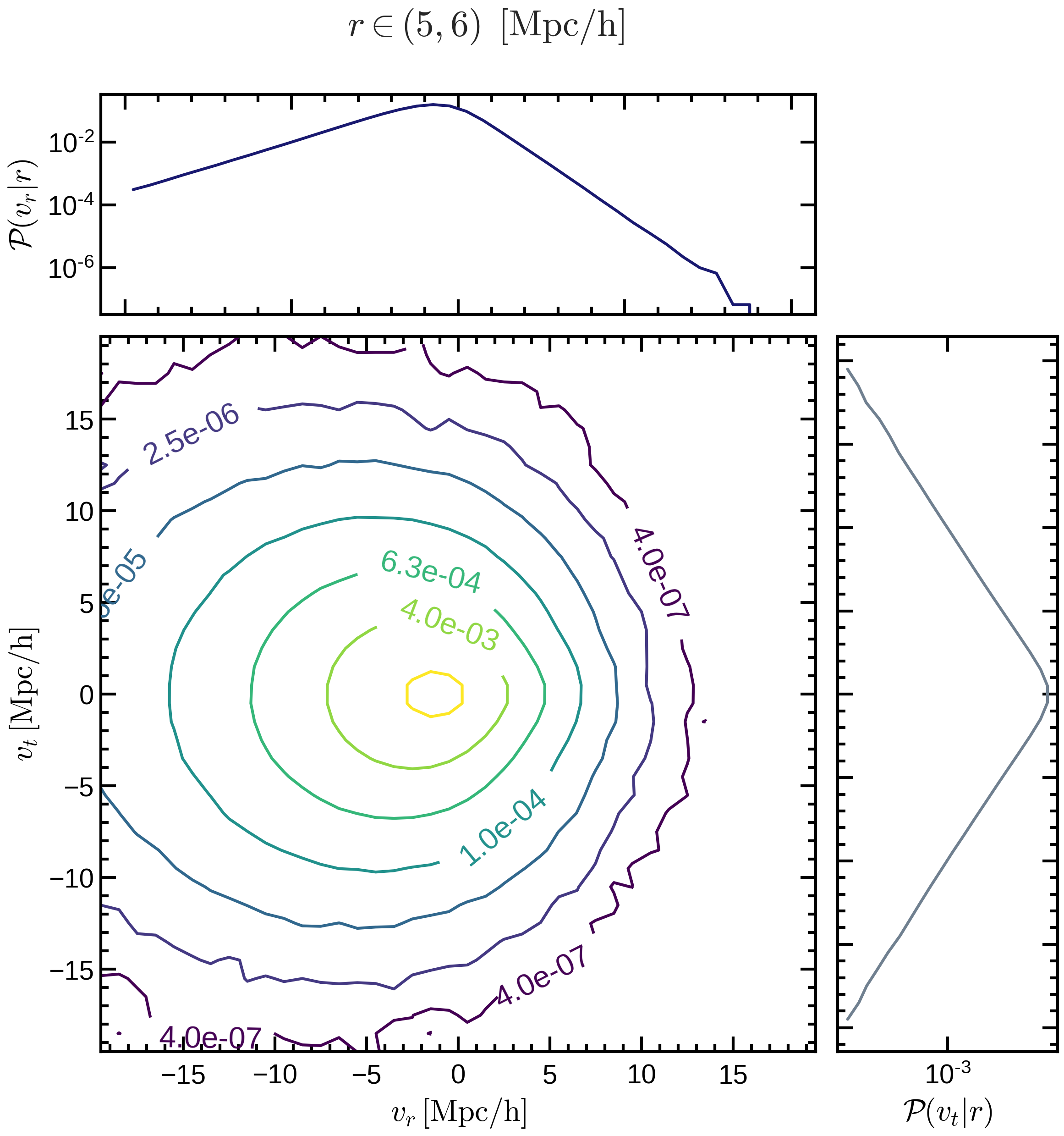}
    \includegraphics[width=0.35\textwidth]{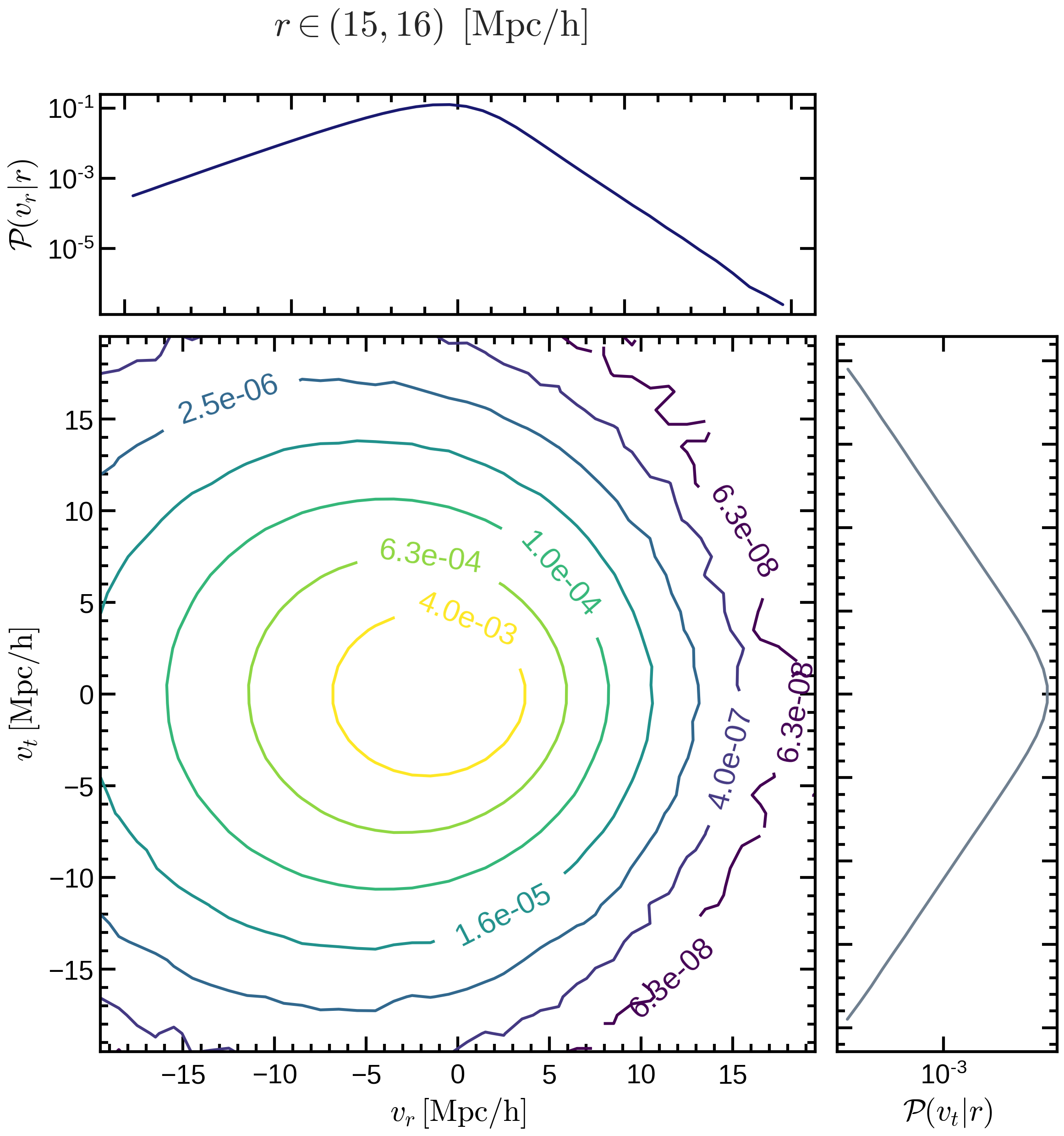}
    }
    \bigskip
    {
    \includegraphics[width=0.35\textwidth]{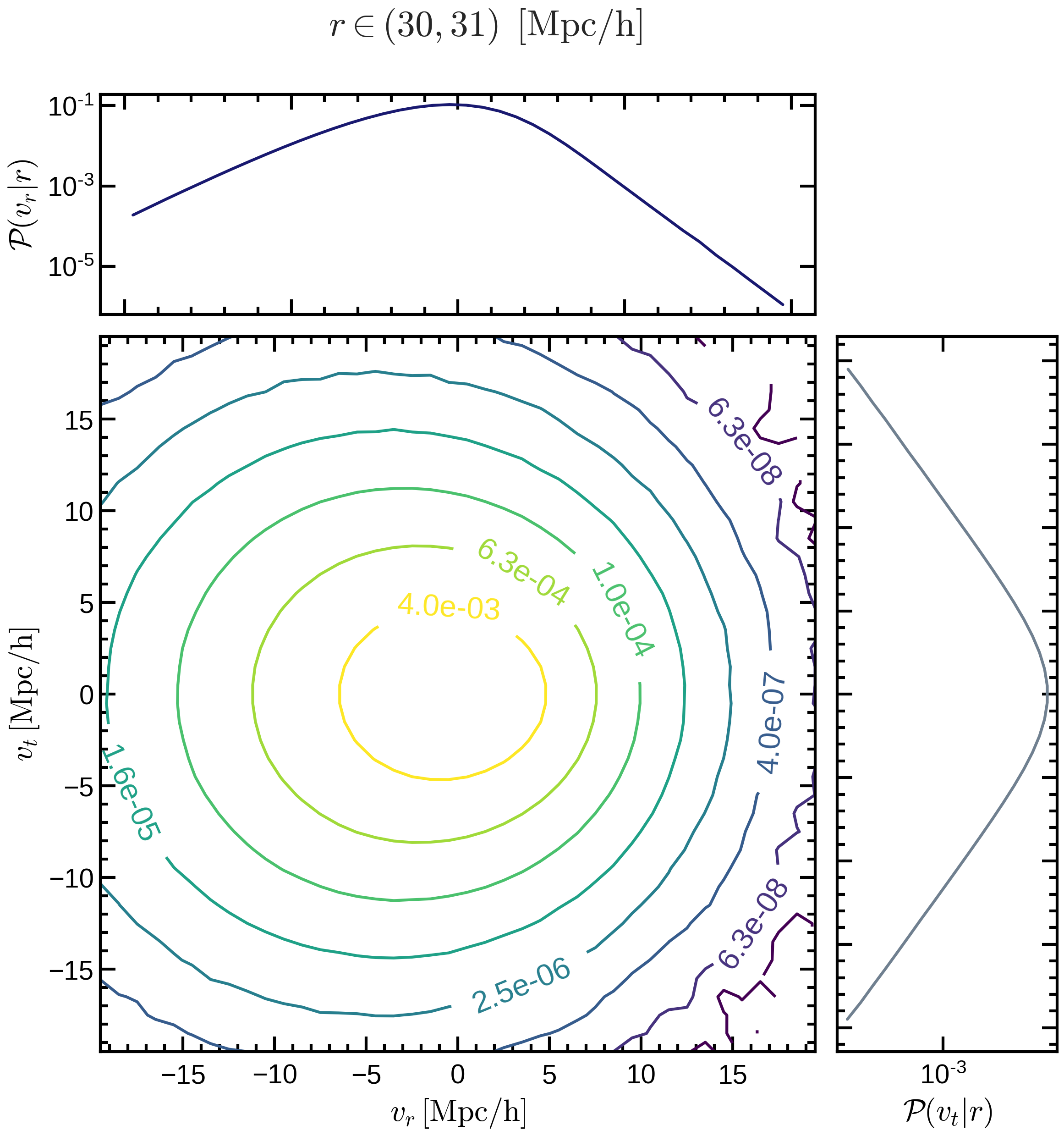}
    \includegraphics[width=0.35\textwidth]{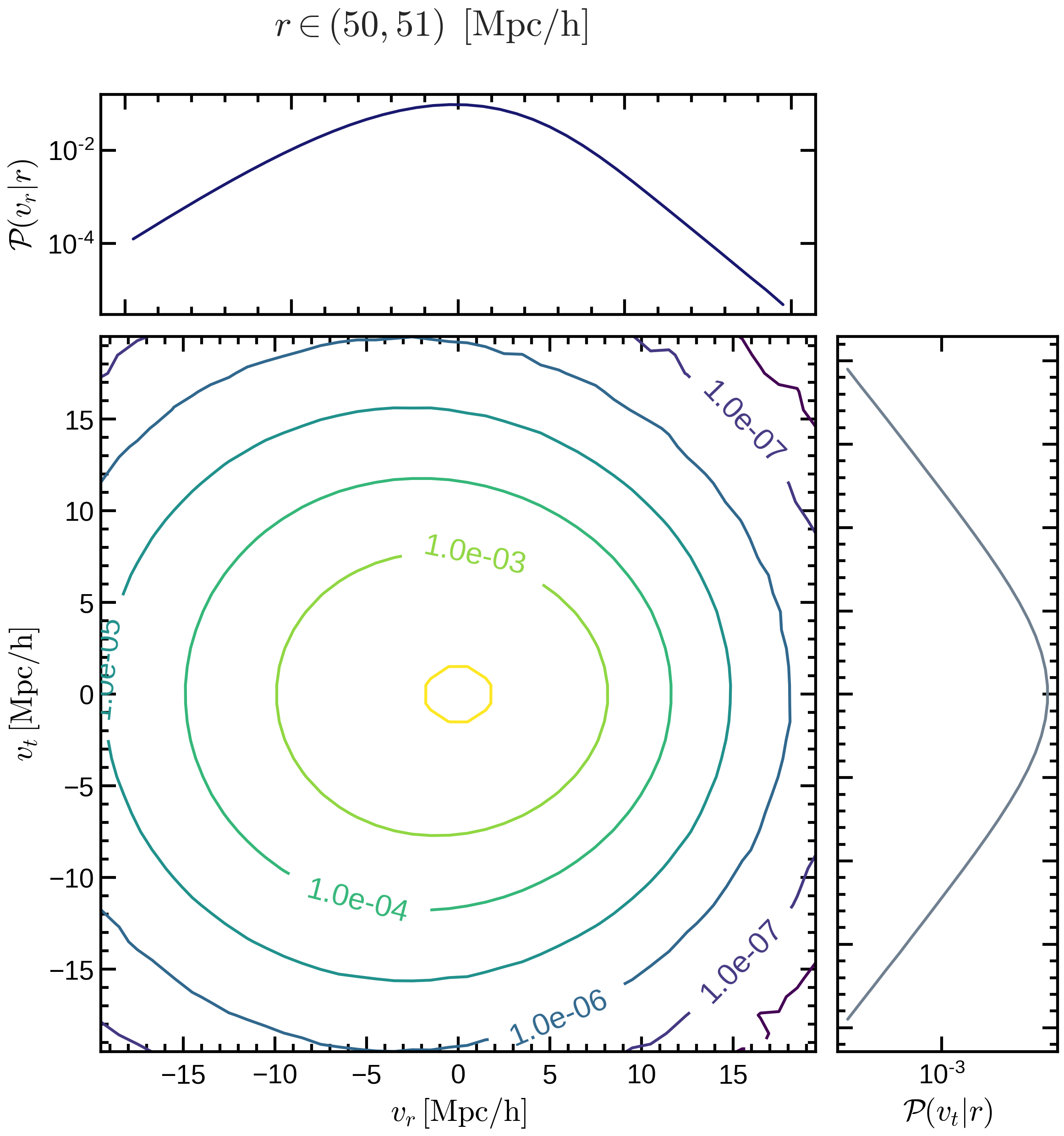}
    }
    \caption{The mean joint probability distribution of the radial and transverse pairwise velocities of dark matter halos measured in N-body simulations. The marginal distributions are shown on the sides. 
    \Takahiro{At small pair separations, }
    infall towards larger structures produces a large skewness in the radial component, and the mean turns more negative. 
    \Takahiro{At large pair separations, }
    the distributions of the two components are symmetric but still show heavy tails.}
    \label{fig:jointpdf}
\end{figure*}

In Fig.~\ref{fig:jointpdf}, we show the radial and transverse pairwise velocity distribution for the halos at different pair separations. Note that the two components are not independent. \Baojiu{This figure shows the same physical picture as Fig.~\ref{fig:line_of_sight_pdf}}. At small pair separations, but larger than halo size, the radial component has a non-zero \Baojiu{(negative)} mean, produced by tracers infalling towards larger objects. Given that the infall velocity is different in different environments \Baojiu{and more pairs are likely to be found in }
high-\Baojiu{density environments where members of a pair tend to approach each other}, the radial distribution is skewed towards negative values. At larger pair separations, the radial skewness 
\Baojiu{becomes smaller}, but it still has heavy tails. Due to \Baojiu{statistical} isotropy, the transverse component is symmetric and has zero mean, although it also shows 
\Carlton{broader tails}
than a Gaussian distribution.

\begin{figure}
\centering
    \includegraphics[width=0.4\textwidth]{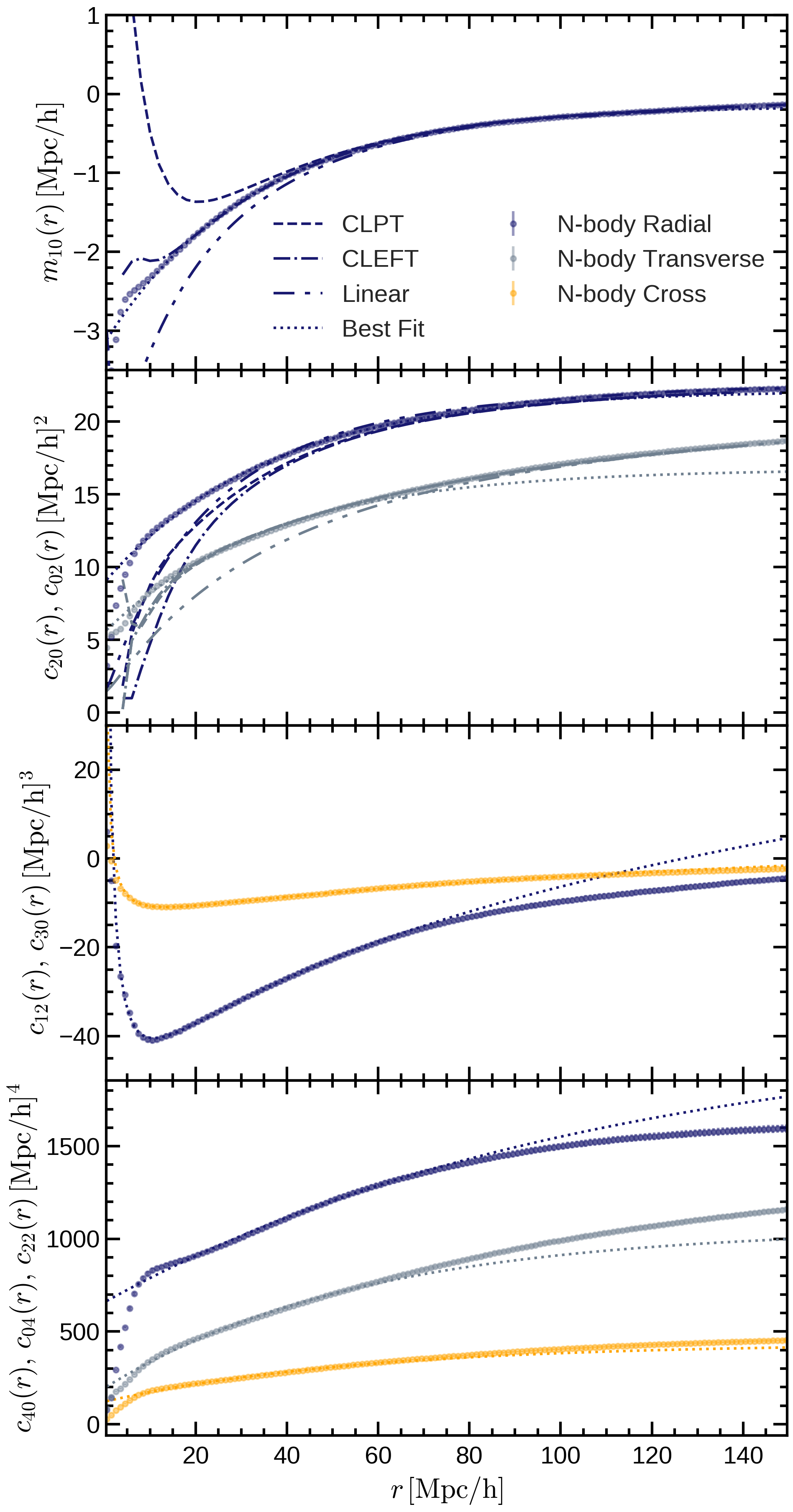}
\caption{
\correction{The four lowest order moments of the radial and transverse pairwise velocity distributions of dark matter halos. In each panel we show the mean measurements from the simulations, together with errorbars showing one standard deviation (note these are too small to be seen). We also show the different perturbation theory predictions for the two lowest order moments. Linear theory is shown in dotted-dashed-dashed lines, CLPT in dashed lines, and CLEFT in dashed dotted lines. Finally, we show the best-fitting curves as dotted lines, which are used to show the accuracy of the Taylor expansion in Section~\ref{sec:taylor}. Note the best-fitting curves have been fitted to the moments on scales smaller than $60\Mpch$. 
}
}

\label{fig:moments}
\end{figure}

The moments of the different distributions are shown in Fig.~\ref{fig:moments}\Baojiu{, where} the definition of the moments is mass-weighted, since the velocity field is only measured where there are tracers, by the number \Baojiu{density} of tracers at a given \Baojiu{separation $r=|{\bf r}|=|{\bf x}_2-{\bf x}_1|$}, 
\begin{equation}
m_{i\, j} = \frac{ \langle (1 + \delta({\bf x}_1))(1+\delta({\bf x}_2)) v_r ^i v_t^j \rangle}{\langle ( 1+ \delta({\bf x}_1))(1 + \delta({\bf x}_2))\rangle},
\label{eq:moment_def}
\end{equation}
where $i$ and $j$ denote the order of the moments in the radial and transverse components respectively. For instance, the radial mean is denoted as $m_{10}$, the second order transverse moment as $m_{02}$, and the third order cross-correlation between the radial and the squared of the transverse component as $m_{12}$. The central moments are analogously defined by
\begin{equation}
c_{i\, j} = \frac{ \langle (1 + \delta({\bf x}_1))(1+\delta({\bf x}_2)) (v_r -m_{10})^i (v_t - m_{01})^j \rangle}{\langle ( 1+ \delta({\bf x}_1))(1 + \delta({\bf x}_2))\rangle}.
\end{equation}
\Baojiu{Statistical} isotropy in the transverse plane implies that only moments with even powers of the transverse component are non-zero. That is $c_{12}$ for the third order moment, and $c_{22}$ for the fourth.

Although \Baojiu{it is not the objective of this paper to develop the} relations between the cosmological parameters and the ingredients of the streaming model (the real-space two-point correlation and the pairwise velocity moments), \Baojiu{for completeness we} show the predictions from different methods \Baojiu{as a summary} of the recent progresses in perturbation theory. 
\Carlton{This exercise}
will show what stage we have reached 
\Carlton{in our efforts to predict these quantities} and what still needs to be done. So far, only predictions for the first two moments of the velocity field have been successfully obtained \Baojiu{from perturbation theory}: 
\begin{itemize}
\item \textit{Linear perturbation theory} -- \cite{1995ApJ...448..494F} shows that the mean pairwise velocity in linear theory is determined by the correlation between the density and velocity fields, $\langle \delta v \rangle$, due to the mass-weighting factors in Eq.~\eqref{eq:moment_def}. The variance, however, is determined by the velocity-velocity coupling \citep{1988ApJ...332L...7G}. In the simplest flavour of Eulerian perturbation theory, there are two free parameters: the linear bias and the growth factor. Higher order corrections to the mean and the variance were computed in \cite{2011MNRAS.417.1913R} by expanding the continuity and Euler equations in powers of the linear density field up to fourth order. They also used a local Lagrangian prescription for the bias \citep{2008PhRvD..78h3519M}, which turned out to be very important to reproduce the real space correlation function, since a local bias in Lagrangian space introduces a non-local bias in Eulerian space \citep{2012PhRvD..86h3540B, 2012PhRvD..85h3509C}.
\item \textit{Convolutional Lagrangian perturbation theory (CLPT)} -- \cite{2014MNRAS.437..588W} extended the formalism of \cite{2013MNRAS.429.1674C}, to include predictions for the lowest-order pairwise velocity moments. The Lagrangian approach formulates the problem in terms of initial positions and displacement field, where the latter fully specifies the motion of the cosmological fluid. Instead of expanding the fluid equations in terms of the linear density field, the expansion is performed on the displacement field that gives the mapping between initial Lagrangian positions and final Eulerian positions. To describe the Lagrangian bias functional, $\delta_h = F[\delta]$, the authors include three free parameters, $b_1$, $b_2$, and $b_s$, which we fit to the real-space two-point correlation function. The first \Carlton{two of these} bias parameters, $b_1$ and $b_2$, are the first and second derivatives of the Lagrangian bias function with respect to \mtrvv{a long-wavelength density contrast,
$\delta^{\rm L}$,}
 whereas $b_s$ encodes the dependence of the bias on \mtrvv{a long-wavelength tidal tensor.}
 The variance of the pairwise velocities is, however, not accurately reproduced by CLPT: a constant shift needs to be added to describe the variance on linear scales. Interestingly, this constant offset is the same for both the radial and transverse components, as one would expect from the effect of virial motions. \Carlton{Including} 
the growth factor, CLPT \Carlton{requires five parameters to describe} clustering in redshift space.
\item \textit{Convolutional Lagrangian effective field theory (CLEFT)} -- \cite{2012JHEP...09..082C} developed an analytical effective field theory to capture the effects of very small scales on large-scale observables. \cite{2016JCAP...12..007V} used this idea, together with CLPT, to predict the lowest-order velocity moments that enter the Gaussian streaming model. They found that predictions for the mean pairwise velocity were greatly improved compared to CLPT, 
\Takahiro{especially}
\Carlton{the}
derivative, which ultimately controls the accuracy of the redshift space quadrupole. Moreover, \Takahiro{it was shown that in the context of effective field theory,} the constant shift in \cite{2014MNRAS.437..588W} was identified as one of the effective parameters to describe the effect of small scales. Increased accuracy comes at the expense of \Carlton{requiring} more free parameters, the effective field theory counter-terms. There are two extra parameters, one for the real space correlation function and 
\Carlton{the other} for the mean pairwise velocity. Therefore, the simplest CLEFT \Baojiu{has} seven parameters.
\end{itemize}

\Baojiu{The top two panels of Fig.~\ref{fig:moments} compare} the predictions for the 
two \Baojiu{lowest-}order moments 
\Baojiu{of} the three different methods\footnote{Perturbation theory predictions have been obtained using the publicly available code \href{https://github.com/martinjameswhite/CLEFT_GSM}{github/CLEFT\_GSM}.}. \Baojiu{The symbols show measurements from simulations.} In the 
\Carlton{upper panel} we show the mean of the radial pairwise velocity. The two extra EFT counter-terms extend the agreement of CLPT with N-body simulations from scales of $\sim60\Mpch$ \Carlton{down} to $\sim20\Mpch$. For the radial and transverse components of the variance, shown in the second panel, \Carlton{the} CLPT and CLEFT predictions are qualitatively similar. The reason 
\Carlton{for this} is that the EFT counter-term is very close to a constant shift in the variance, which is already included in CLPT to match N-body simulation results. The moment predicted with the lowest accuracy is the radial component of the variance, where per cent-\Carlton{level} predictions are limited to scales above $40\Mpch$. Note that the radial component of the variance, $c_{20} = m_{20} - m_{10}^2$, 
\Carlton{has} a contribution from the mean pairwise velocity and will also be affected by errors \Carlton{in} modelling non-linear infall.

\correction{\Baojiu{The bottom two panels of Fig.~\ref{fig:moments} show the simulation measurements of the third- and fourth-order moments (symbols).} Perturbation theory predictions for moments higher than the second have only been obtained for the third order moment using CLPT in \cite{2015PhRvD..92f3004U}. However, the authors found that it fails to capture the non-Gaussian effects encoded in the skewness for scales below $100 \Mpch$. Since third and fourth order moments only play a role on the accuracy of the redshift space correlation function on small scales, it is extremely difficult to produce accurate enough predictions to unlock access to the cosmological information contained on those scales. To improve these predictions, we plan to explore both effective field theory extensions to CLPT, and the use of emulators for the moments on small scales.}

\begin{figure}
\centering
    \includegraphics[width=0.4\textwidth]{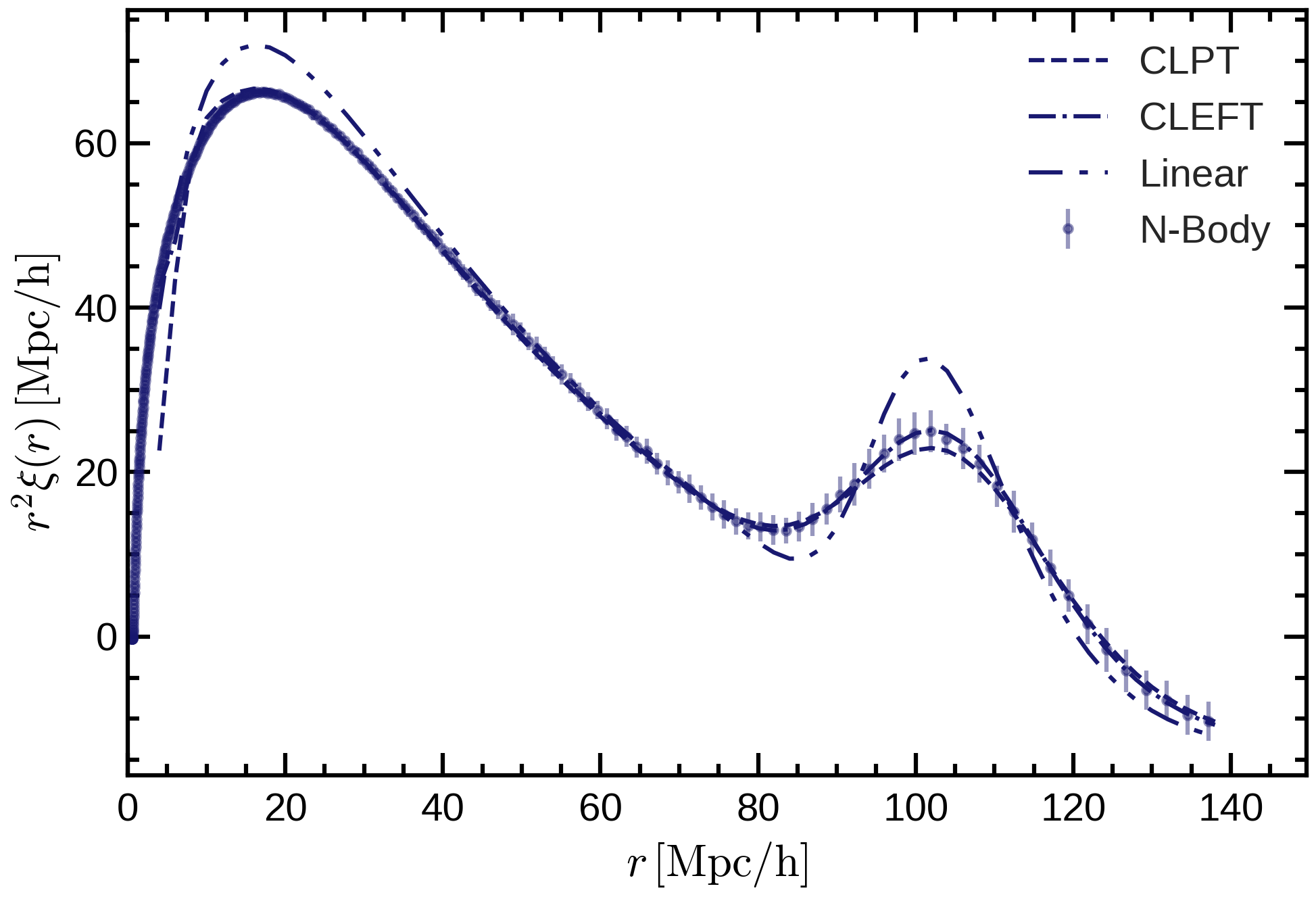}
\caption{The 
real space correlation function 
\Carlton{measured from the simulations for dark matter halos}, compared with predictions from CLPT and CLEFT. 
Both \Carlton{these perturbation theory} predictions use a Lagrangian prescription for the bias, and have been computed 
\Carlton{by simultaneously fitting} 
the correlation function and the 
\Carlton{two lowest order} 
pairwise velocity moments. For more details on the fitting see  \Baojiu{Appendix} \ref{app:pt_details}.}
\label{fig:tpcf}
\end{figure}

In Fig.~\ref{fig:tpcf}, we also show \Baojiu{simulation measurements (symbols)} of the real-space two-point correlation function \Carlton{of dark matter halos}, together with the predictions 
\Carlton{using} both CLPT \Baojiu{(dashed line)} and CLEFT \Baojiu{(dash-dotted line)}. The CLEFT prediction is  accurate over a broad range of scales \Baojiu{-- it gives per cent-accuracy results on scales between $10$ and $70\Mpch$ --} at the expense of only one extra free parameter. For more details on the accuracy of the different perturbation theory models, we refer the reader to \Baojiu{Appendix} \ref{app:pt_details}.

\subsection{Fitting the pairwise velocity distribution}
\label{sec:parameters}

To infer the parameters of the \Baojiu{Gaussian and ST} distributions that best fit the simulation \Baojiu{measurements}, we 
\Baojiu{use} two different methods,
\begin{itemize}
    \item \textit{Maximum likelihood estimation}, found by maximising the probability that the model 
    \mtrv{reproduces}
    the 
    \Carlton{simulation measurements.} 
    \Carlton{This} is equivalent to a least-$\chi^2$ 
    fit to \Baojiu{the simulation measurements using the given Gaussian or ST distribution function} when the errors are approximately Gaussian distributed. We refer to this method as `ML' occasionally in this paper. The $\chi^2$ we minimise is,
\begin{equation}
    \chi^2(\mathbf{r}) = \sum_{v_\parallel}\frac{ \left( \mathcal{P}_{\mathrm{measured}}(v_\parallel | \mathbf{r}) - \mathcal{P}_{\mathrm{model}}(v_\parallel | \mathbf{r})  \right)^2}{\mathcal{P}_{\mathrm{model}}(v_\parallel | \mathbf{r}) },
\end{equation} where $\mathcal{P}_{\mathrm{model}}$ is either a Gaussian or an ST distribution.
    \item \textit{Method of moments}, that uses the analytical relation between the parameters of the distribution and its 
    \Baojiu{lowest-}order moments to convert the 
    moments \Carlton{estimated} from the \Baojiu{simulation} \Carlton{measurements} 
    into distribution parameters. 
\end{itemize}
If the 
distribution \Carlton{measured from the simulation} and the 
fitted distribution are the same, both methods are equivalent. However, this is not the case when the fitted distribution is an approximation to the 
\Carlton{simulation results} or when noise is present. 

In Fig.~\ref{fig:line_of_sight_pdf} we show the 
\Carlton{best-fitting}
distributions for both the Gaussian and the ST models using these two approaches. For the Gaussian case the conversion between moments and parameters is trivial, while for the ST model we \Baojiu{have used} the relations given in \Baojiu{Appendix}~\ref{app:moments_st} to obtain the model parameters given the four \Baojiu{lowest-}order moments. In 
this figure we can see that even 
\Baojiu{for large pair separations} the Gaussian approximation is inaccurate, \Baojiu{where} the method of moments and the maximum likelihood estimation produce slightly different results, both being 
\Baojiu{poor} approximations. 

The ST model, however, is flexible enough to represent the varying degrees of skewness and kurtosis over a broad range of \Baojiu{pair separations} when using the method of moments. At large {separations}, the maximum likelihood estimate and the method of moments produce similar distributions. Nonetheless, on small scales the tails of the distribution are mis-estimated by the maximum likelihood method.

\correction{For a more detailed comparison of the different models around the peak of the distribution see Fig.~\ref{fig:line_of_sight_pdf_nolog}.}

\subsection{The redshift space correlation function}
\label{sec:redshift_space}
In this subsection we use the Gaussian and ST models of the pairwise velocity distribution with the streaming model (Eq.~\ref{eq:streaming}) to predict redshift space clustering. 
\Baojiu{We will focus on} the mapping between real and redshift spaces, 
\Baojiu{and show that using the more flexible ST distribution for pairwise velocity leads to more accurate predictions of the higher order  redshift-space multipoles than are obtained with the simpler GSM model. For this reason,} we measure all the real-space quantities from the simulation, \Baojiu{including} the \Baojiu{real-space} halo two-point correlation function and the pairwise velocity \Baojiu{distribution} moments, \Carlton{ as inputs} to reproduce the redshift space clustering by using Eq.~\eqref{eq:streaming}. \Baojiu{The impact of the accuracy of the modelling of the individual ingredients of the streaming model will be studied in a later section.}

The pairwise velocity distribution has been measured in the range 
\mtrvv{$0 <r_\parallel/[\Mpch]<70$ and $0<r_\perp/[\Mpch]<50$ in bins spaced by $0.5~\Mpch$.} To perform the streaming model integration in Eq.~\eqref{eq:streaming} we have used the Simpsons rule, with a linear interpolation of the real space correlation function and the pairwise velocity distribution. 

Due to the difficulty of analysing two dimensional plots, together with the complex covariance matrix between the different 
\Carlton{measurements} for $\xi^S(s_\perp, s_\parallel)$, 
it is common to decompose the redshift space correlation function into multipole moments using its Legendre expansion \citep{1998ASSL..231..185H},
\begin{equation}
    \label{eq:legendre_exp}
    \xi(s, \mu) = \sum_\ell\xi_\ell(\mu) L_\ell(\mu),
\end{equation}
where $\ell$ is the order of the multipole \Baojiu{and $L_\ell(\mu)$ is the Legendre polynomial at the $\ell$-th order, which depends on the angular coordinate $\mu=\cos\theta$}. The redshift space correlation function is symmetric in $\mu$, so only even values of $\ell$ give a non-zero contribution. Inverting~Eq.~(\ref{eq:legendre_exp}), we find that the multipole moments are given by
\begin{equation}
    \xi_\ell(s) = \frac{2 \ell + 1}{2} \int_{-1}^1 \xi(s, \mu) L_\ell(\mu) d\mu.
\end{equation}
The three lowest multipoles are denoted as monopole ($\ell = 0$), quadrupole ($\ell = 2$) and hexadecapole ($\ell = 4)$.  Recent cosmological analyses are based mainly on the monopole and quadrupole moments, however the cosmological information carried by the hexadecapole has also been shown to be important \citep{2011PhRvD..83j3527T}.

\begin{figure}
\centering
    \includegraphics[width=0.4\textwidth]{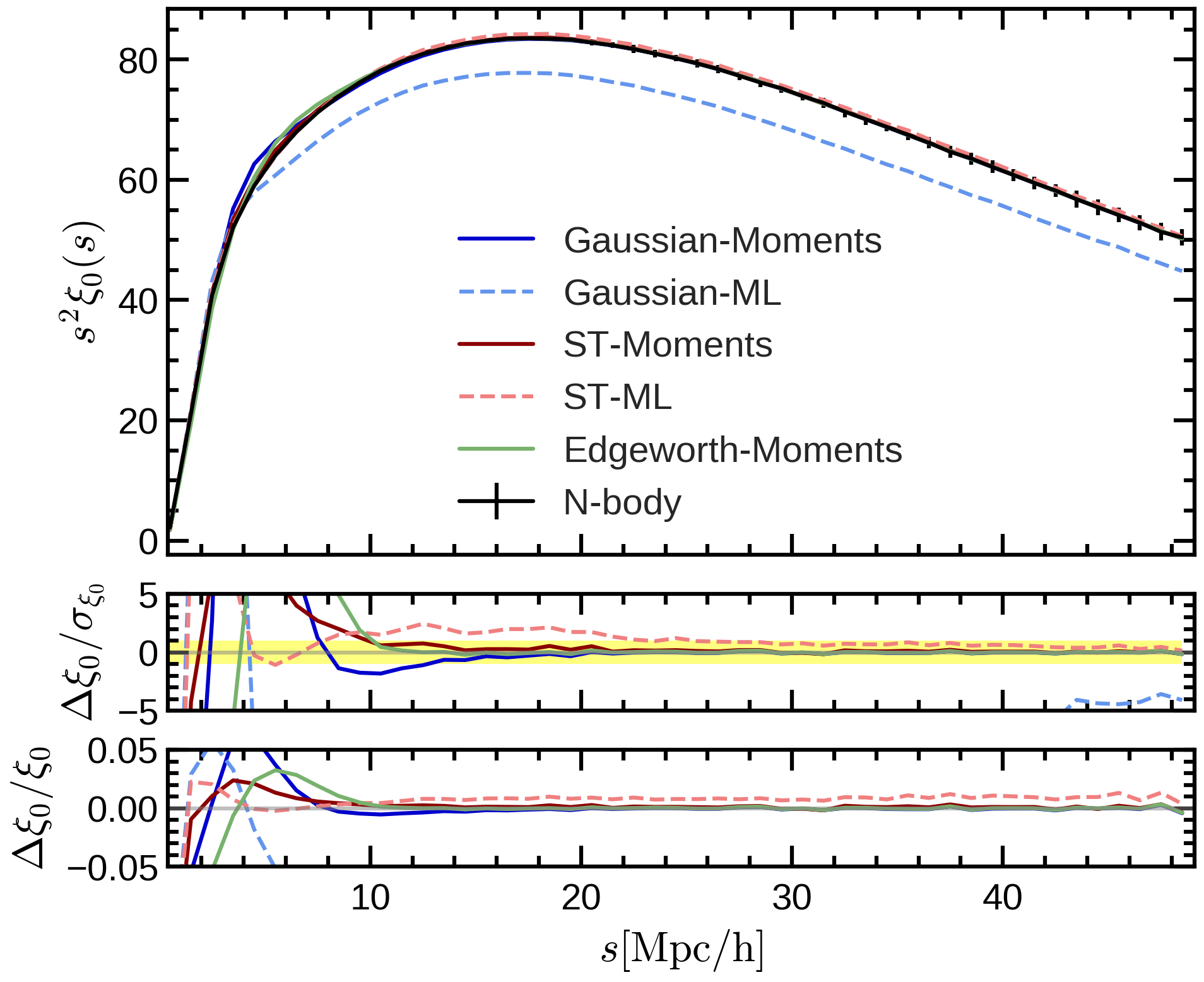}
    \includegraphics[width=0.4\textwidth]{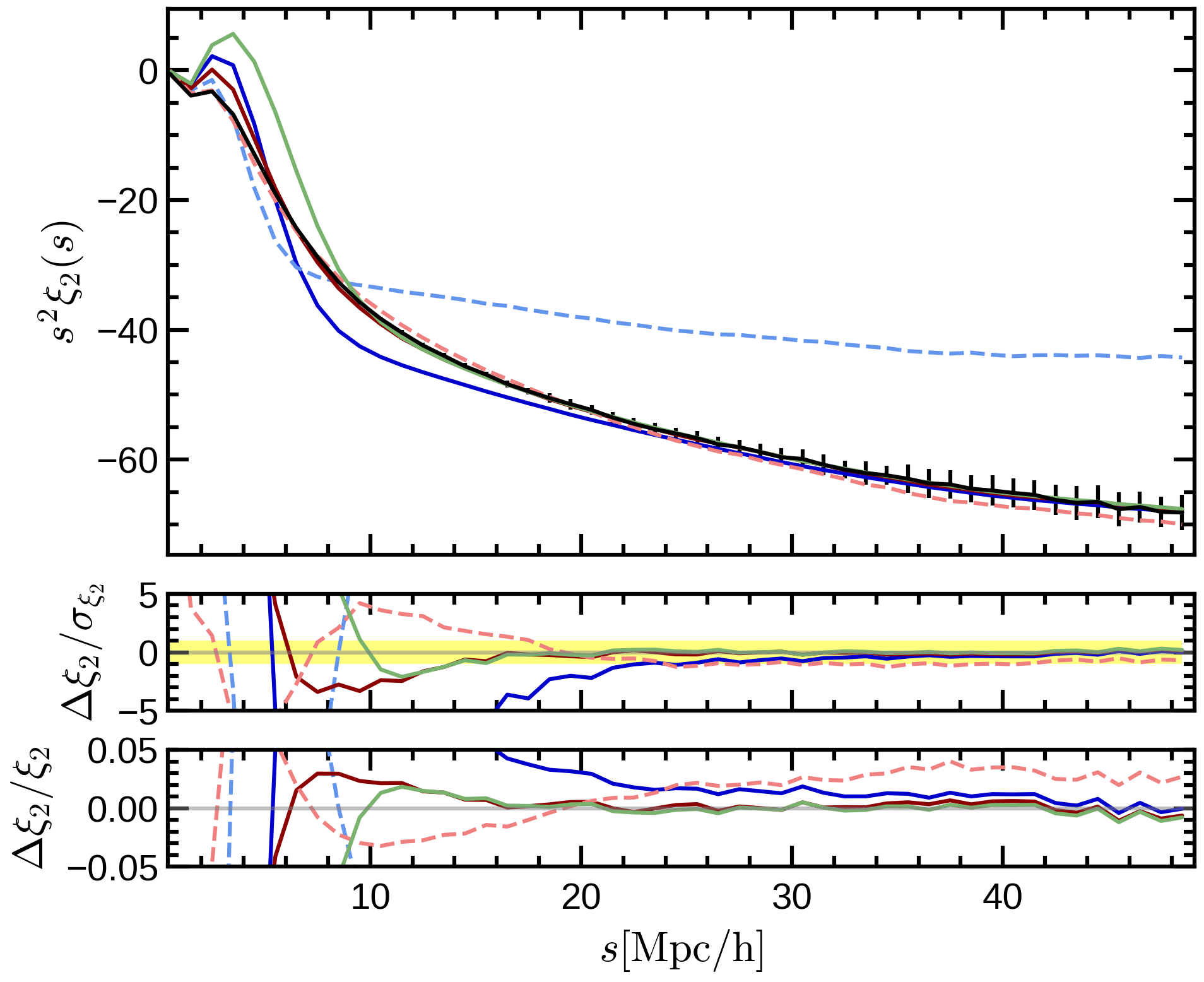}
    \includegraphics[width=0.4\textwidth]{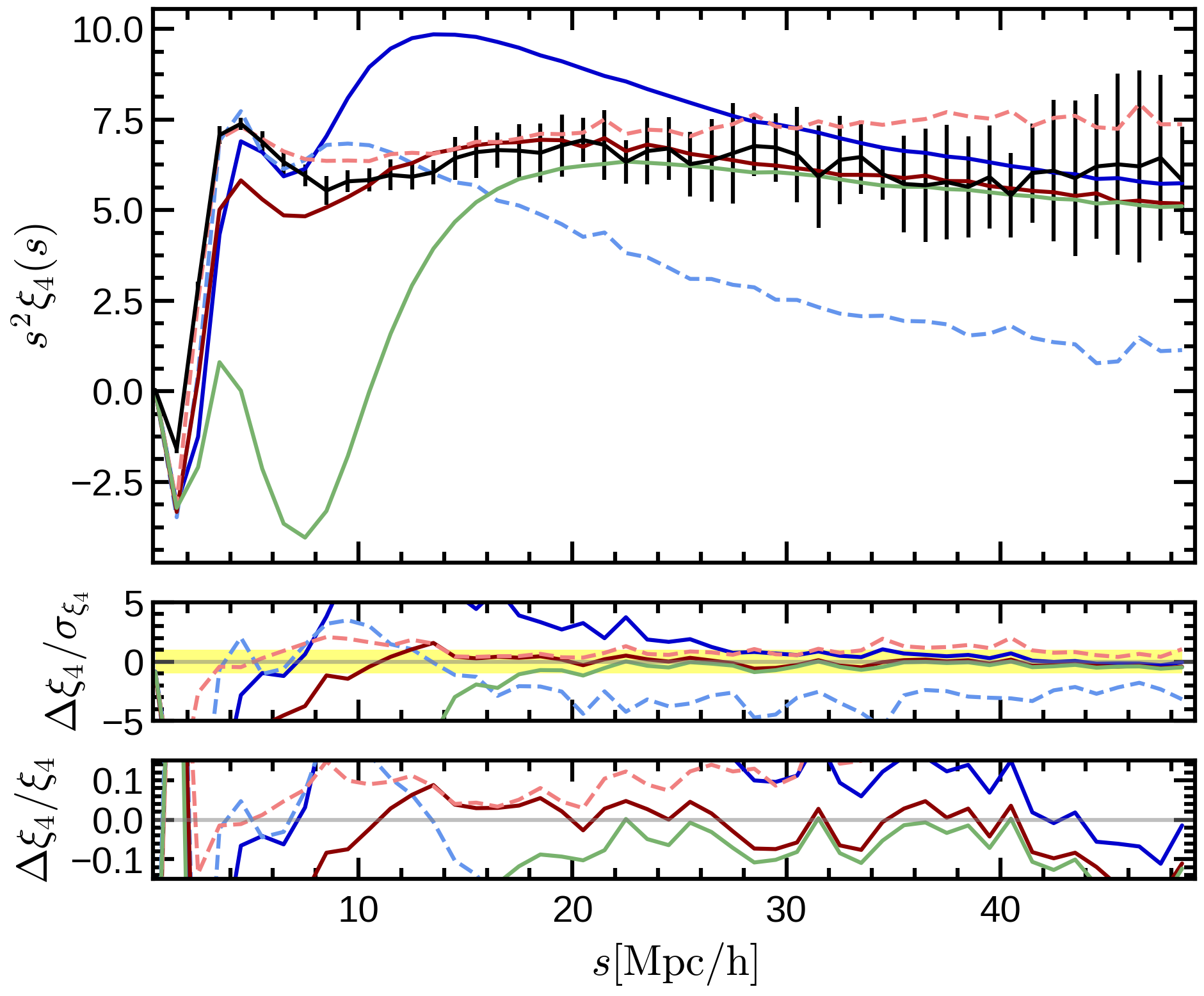}
\caption{Comparison of the accuracy of the different models \Carlton{for} reproducing the multipoles of the 
\Carlton{redshift} space correlation function. In the upper sub-panels the \correction{ multipole directly measured from the simulation} is shown together with the model predictions. In the lower sub-panels the deviation between the model and the simulation in units of the variance calculated across the different independent simulations are shown. The yellow bands show the $1\sigma$ deviation.}
\label{fig:multipoles}
\end{figure}

We show these three multipole moments predicted by the different models, \Baojiu{as well as the measurements from the simulations,} in Fig.~\ref{fig:multipoles}. \Baojiu{In the lower subpanels of each panel we show the relative differences between the model predictions and the simulation results in units of the standard deviation ($\sigma$) (middle subpanel) calculated using the 15 simulation realisations \mtrvv{each of which has a volume of $8~(h^{-1}{\rm Gpc})^3$}, \correction{and the relative percent error in the lowest subpanel}. The yellow horizontal shaded bands represent the $\pm1\sigma$ ranges on the multipoles.} 

Surprisingly, the two Gaussian distributions that we found by using the method of moments and the maximum likelihood estimate 
\Carlton{yield} 
multipoles that can be more than five standard deviations away from each other. Furthermore, the Gaussian distribution obtained using the method of moments reproduces the three multipoles within one standard deviation for scales larger than approximately
$30\Mpch$, although it gives a very poor fit to the pairwise velocity distribution on these scales; \Baojiu{cf.~Fig.~\ref{fig:line_of_sight_pdf}.} 

Regarding the ST model, although the maximum likelihood ST lies closer than the Gaussian method of moments to the pairwise velocity measured in the simulation (see 
\Baojiu{Fig.~\ref{fig:line_of_sight_pdf})}, it gives a biased result for the multipoles. \Baojiu{On the other hand,} the ST model found by the method of moments is able to reproduce the correct clustering down to scales of around $10\Mpch$. 

\correction{The Edgeworth model does improve the predictions of the multipoles compared to the Gaussian Streaming Model, however to extend its validity to even smaller scales we need to also add fourth order moments.}

Note that, as a result of the large simulated volume, the error bars on the monopole and quadrupole on scales below $20\Mpch$ are extremely small, meaning that the one sigma 
\Baojiu{deviations} for the monopole and quadrupole \Baojiu{(the yellow horizontal bands in the lower subpanels of Fig.~\ref{fig:multipoles})} are within one per cent of the mean measurement up to $20\Mpch$.

Finally, although the measurement of the hexadecapole is itself very noisy, the ST model is within one standard deviation for scales larger than around $10\Mpch$, whilst the Gaussian model on those scales is already more than five sigma away from the measurement from the simulations.

To sum up, we \Carlton{have} found that the use of the method of moments is critical to accurately reproduce the clustering on quasi-linear scales. The accuracy of the Gaussian streaming model we obtain is consistent with previous findings \citep{2011MNRAS.417.1913R, 2014MNRAS.437..588W,2016MNRAS.463.3783B}: the prediction is within the measurement errors from the simulations for scales larger than $30\Mpch$. However, the model prediction 
\Carlton{rapidly diverges from the simulation results} 
on smaller scales. On the contrary, the ST model is able to reproduce the redshift space clustering very 
accurately on scales 
\Baojiu{down to} $10\Mpch$, by introducing a \Baojiu{pairwise velocity} distribution that incorporates the skewness and kurtosis of the pairwise velocity PDF.

On the other hand, we need to understand why the Gaussian model reproduces the clustering on scales above $30\Mpch$ more accurately than the ST distribution obtained through the ML method, 
\Baojiu{even though the latter is a better description of} the pairwise velocity distribution on those scales, as shown in Fig.~\ref{fig:line_of_sight_pdf}. 
\Baojiu{To this end}, we will study the behaviour of the integrand of Eq.~\eqref{eq:streaming} \Baojiu{in more detail in the next section}.


\section{The importance of the moments for accurate  clustering predictions}

In this section we 
show how the accuracy of the streaming model on quasi-linear scales is directly related to the lowest order moments of the pairwise velocity distribution. We start by studying how well the different models reproduce the streaming model integrand.

\subsection{Lessons from the streaming model integrand}

\begin{figure}
\centering
    \includegraphics[width=0.4\textwidth]{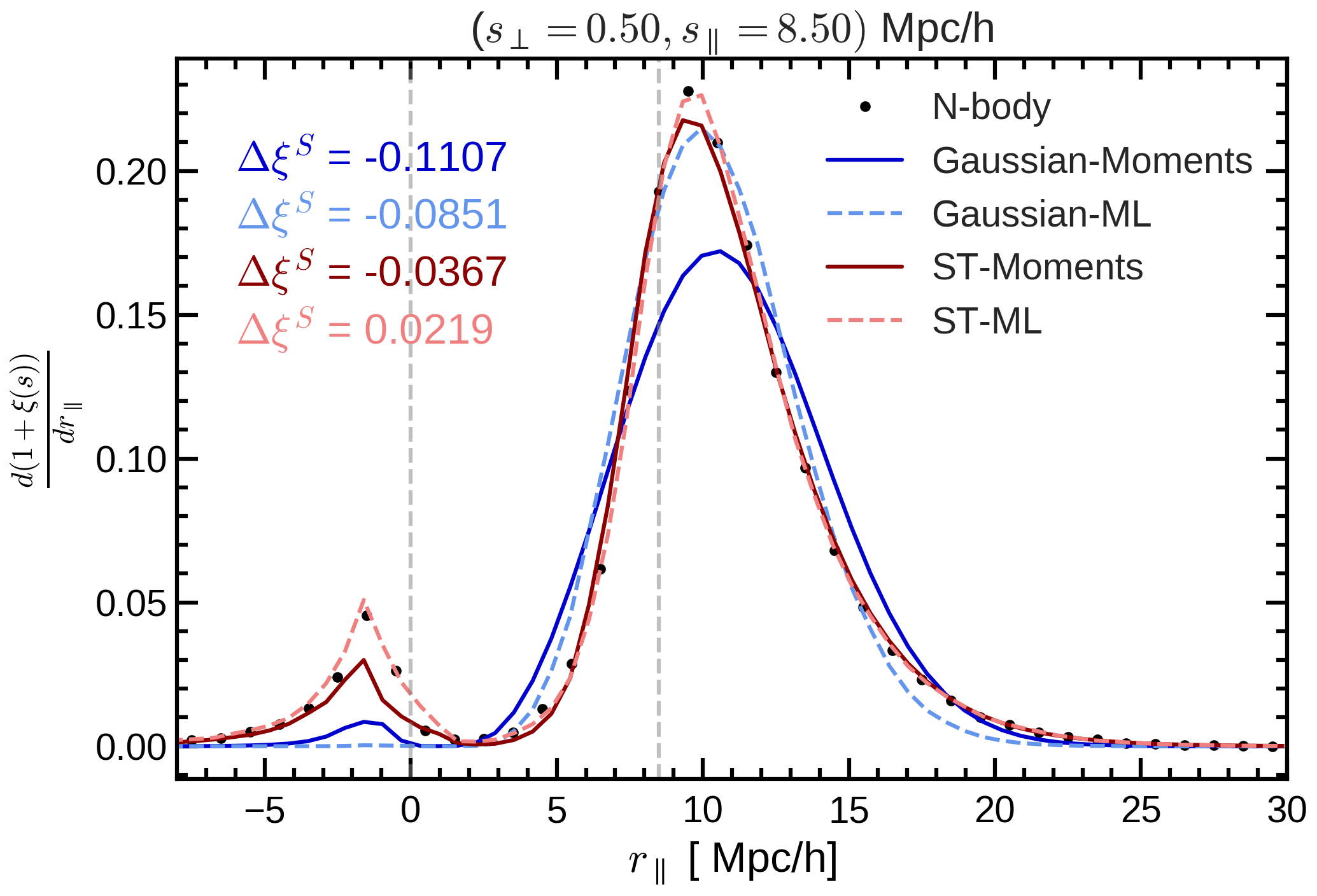}
    \includegraphics[width=0.4\textwidth]{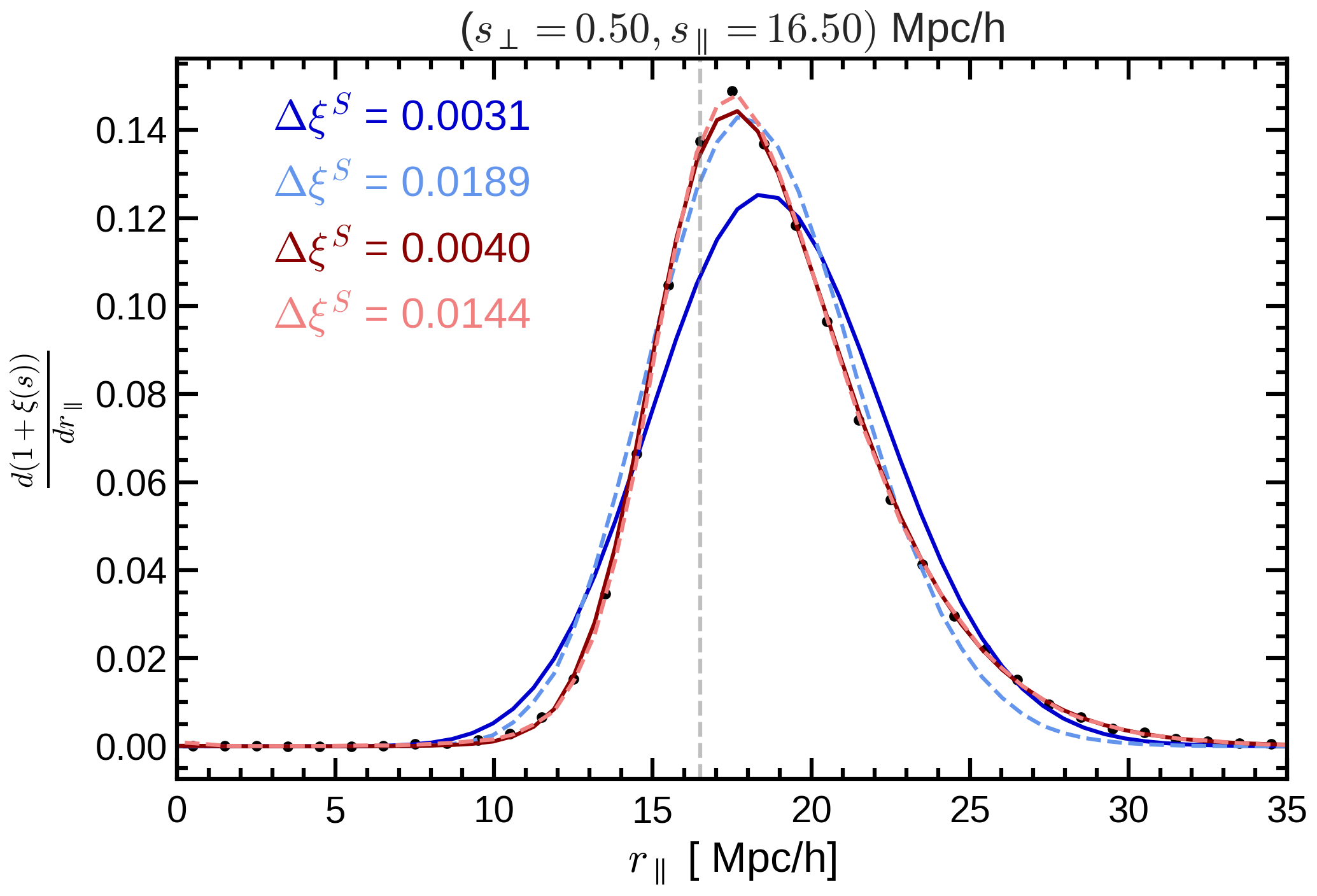}
    \includegraphics[width=0.4\textwidth]{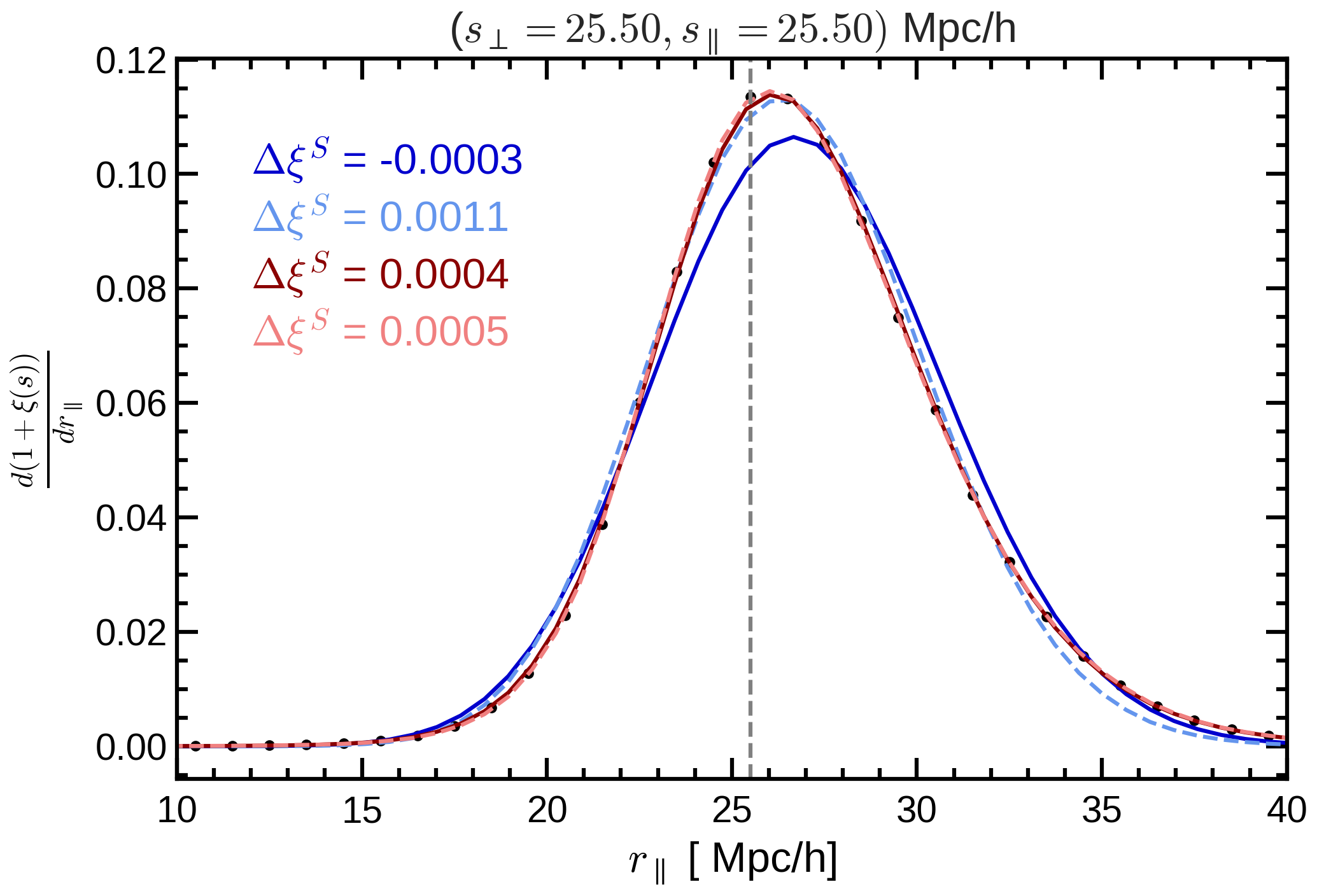}
\caption{Integrand of Eq.~\ref{eq:streaming} shown for different redshift space pair separations. At small pair separations and for small $\mu$ (top panel) we find two peaks situated at $r_\parallel = 0$ and $r_\parallel = s_\parallel$, marked by the grey vertical dashed lines. For larger $\mu$ (middle and bottom panels), the second peak dominates since the correlation function decays rapidly at large separations. The result of the integral for the different models minus the integral obtained using the 
pairwise velocity distribution 
\Carlton{measured from} 
the simulations, $\Delta \xi(s)$, is plotted with different models shown \Carlton{by the different colours and line styles as shown by the legend in the top panel.} 
}
\label{fig:integrand}
\end{figure}

We show the integrand of Eq.~\eqref{eq:streaming}, for a few pair separation vectors, in Fig.~\ref{fig:integrand}. In broad terms the integrand is the outcome of a competition between the probability of finding a pair of haloes at a given 
\Carlton{separation}, i.e. the two-point correlation function, and the probability that the pair has the necessary relative velocity to move from real space position $\mathbf{r}$ to redshift space position $\mathbf{s}$. Whilst the first quantity is evaluated as $\xi\left(\sqrt{s_\perp^2 + r_\parallel^2}\right)$ \Baojiu{for fixed $s_\perp$}, and therefore peaks at $r_\parallel \sim 0$, the latter is evaluated as $\mathcal{P}(v_\parallel = s_\parallel - r_\parallel)$, and peaks around its mean, close to $v_\parallel \approx 0$  ($r_\parallel \approx s_\parallel$) \Baojiu{for large pair separations.} 
 
\Baojiu{The effect of this competition can be seen in Fig.~\ref{fig:integrand}. For large pair separations, e.g., as shown by the middle and bottom panels, the real-space correlation function is small, $\xi\left(\sqrt{s_\perp^2 + r_\parallel^2}\right)\ll1$, so that the integrand is dominated by $\mathcal{P}(v_\parallel = s_\parallel - r_\parallel)$ and has a peak at $r_\parallel \approx s_\parallel$. On the other hand, for small pair separations (the top panel), $\xi$ is no longer negligible and the integrand acquires a second, albeit smaller peak around $r_\parallel\approx0$.}

As for the different streaming models, the Gaussian one obtained through the method of moments systematically shifts the \Baojiu{main} peak of the integrand \Baojiu{from its true position}, and makes it wider. Although this seems to be a poorer estimate of the integrand than the Gaussian model obtained through maximising the likelihood, \Baojiu{which is consistent with what Fig.~\ref{fig:line_of_sight_pdf} suggests,} it \Carlton{predicts the clustering multipoles with a precision that is one order of magnitude higher} after integrating, as can be seen on the resulting redshift space correlation function annotated on Fig.~\ref{fig:integrand}. This same effect is present on all scales larger than $s \approx 30\Mpch$.

\Baojiu{More interestingly, both the ST moments and the ST ML methods give visually much better predictions for the integrand than the Gaussian moments method, which is a consequence of the pairwise velocity distribution being non-Gaussian for all pair separations. However, as shown in the previous section, after integration} we find that 
the Gaussian model 
\Carlton{yields} a comparable accuracy to 
the non-Gaussian ST model on scales larger than $30\Mpch$ for the monopole and quadrupole. This
\Baojiu{coincidental} behaviour has been noted previously by \cite{2018MNRAS.479.2256K}.  
Taking the middle panel of Fig.~\ref{fig:integrand} as an example, the ``errors'' of the integration in Eq.~\eqref{eq:streaming}, $\Delta\xi^S$, defined as the difference between the integration of the model curve and the integration using the simulation results (black dots), for the four streaming models considered here, are shown in the figure labels. We note that the Gaussian moments method gives a slightly smaller error than the ST Moments at the particular pair separation shown. As the former underestimates the integrand for $14\lesssim{r}_\parallel/(\Mpch)\lesssim20$ and $r_\parallel\gtrsim26\,\Mpch$, and overestimates it in other regimes, this seems to suggest that a precise cancellation of the errors from different $r_\parallel$ intervals takes place, which makes the final integration result accurate. However, this cancellation of errors happens for all larger pair separations $s > 30 \, \Mpch$. In the next subsection, we will show \Baojiu{that this is a consequence of the integration being sensitive only to the moments of the pairwise velocity distribution. In particular, for large pair separations it is the} 
two \Baojiu{lowest-}order moments which dominate the outcome of the integral Eq.~\eqref{eq:streaming}, 
while higher order moments only become important on scales smaller than $30\Mpch$ (Fig.~\ref{fig:multipoles}).
\subsection{The importance of the moments on quasi-linear scales}
\label{sec:taylor}

\begin{figure}
    \centering
    \includegraphics[width=0.4\textwidth]{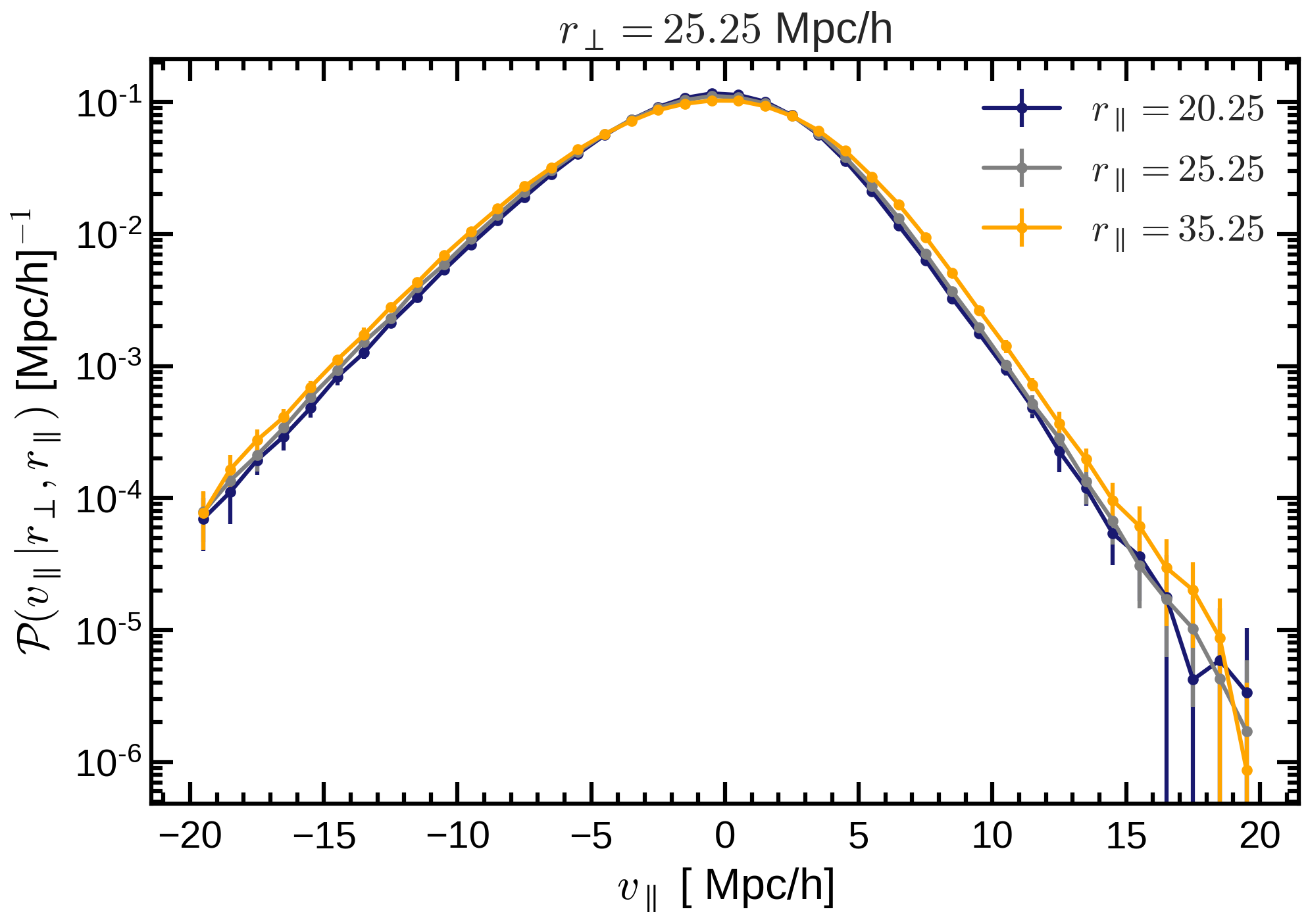}
    \caption{The dependence of the pairwise velocity distribution on $r_\parallel$, for fixed $r_\perp=25.25\Mpch$ and for three values of $r_\parallel$ 
    \Carlton{over a range of} $15\Mpch$ ($20.25, 25.25$ and $35.25\Mpch$). The distribution 
    has \Carlton{only} a weak dependence on $r_\parallel$, which is why the Taylor expansion described in the text works.}
    \label{fig:slowly_varying}
\end{figure}

The integration in the streaming model, Eq.~\eqref{eq:streaming}, is different from taking the expectation value of $1+\xi^R(r)$ since the pairwise velocity distribution $\mathcal{P}(v_\parallel|{\bf r})$ is different for different $r_\parallel$ values, rather than a fixed probability distribution function $\mathcal{P}(v_\parallel)$. However, $\mathcal{P}(v_\parallel|{\bf r})$ is a slowly varying function of pair separation $r_\parallel$, for $r_\parallel \gtrsim15\Mpch$, as can be seen in Fig.~\ref{fig:slowly_varying}. The outcome of the streaming model integral for the values $(s_\perp = r_\perp = 25.5, s_\parallel = 25.5)$, shown in the bottom panel 
of Fig.~\ref{fig:integrand}, 
\Carlton{ is dominated by contributions} from the pairwise velocity distribution in the range $20 \, \Mpch < r_\parallel < 35 \, \Mpch$, which is the range of values shown in Fig.~\ref{fig:slowly_varying}. The same features is found at 
\Carlton{other} separations larger than about $10\, \Mpch$.

Therefore, we can Taylor expand the integrand around its peak at $r_\parallel = s_\parallel$ as follows, 
\begin{equation}
\label{eq:expansion_integrand}
\begin{split}
    \left(1 + \xi^R(r) \right) \mathcal{P}(v_\parallel | \mathbf{r}) & \approx   \left(1 + \xi^R(s) \right) \mathcal{P}(v_\parallel | \mathbf{s})\,\,+ \\ + & \sum_n \frac{1}{n!} (r_\parallel - s_\parallel)^n\frac{d ^n} {d s_\parallel^n}\left((1 + \xi^R(s))  \mathcal{P}(v_\parallel | \mathbf{s})  \right) .
\end{split}
\end{equation}

This expansion was already used by \citet{1995ApJ...448..494F}, \citet{2004PhRvD..70h3007S} and \citet{2015MNRAS.446...75B} to obtain the Kaiser limit of the streaming model. Here, we will show that this is still accurate on quasi-linear scales.

Inserting Eq.~\eqref{eq:expansion_integrand} into Eq.~\eqref{eq:streaming}, we find that the derivatives with respect to $s_\parallel$ can be taken out of the integral over $r_\parallel$, together with the real space correlation function, and \Baojiu{therefore after integration} we are left with the derivatives of the moments through the dependency of $v_\parallel$ on $r_\parallel$. For the lowest order term on the right-hand side of Eq.~\eqref{eq:expansion_integrand}, we find after a change of variables $v_\parallel = s_\parallel - r_\parallel$,
\begin{equation}
    \int_{-\infty}^{\infty} d v_\parallel \left(1 + \xi^R(s) \right) \mathcal{P}(v_\parallel | \mathbf{s}) =  \left(1 + \xi^R(s) \right),
\end{equation}
whilst for the higher order terms,
\begin{equation}
\begin{split}
    \int_{-\infty}^{\infty} d v_\parallel (-1)^n  \sum_n \frac{1}{n!} v_\parallel^n  \frac{d^n}{d s_\parallel^n}  \left((1 + \xi^R(s))  \mathcal{P}(v_\parallel | \mathbf{s})  \right)= \\
     \sum_n \frac{(-1)^n}{n!} \frac{d^n}{d s_\parallel^n}
    \left((1 + \xi^R(s)) \int_{-\infty}^{\infty} d v_\parallel  v_\parallel^n \mathcal{P}(v_\parallel | \mathbf{s}) \right) = \\
         \sum_n \frac{(-1)^n}{n!}  \frac{d^n}{d s_\parallel^n} \left((1 + \xi^R(s)) m_n(\mathbf{s}) \right),
\end{split}
\end{equation}
where $m_n$ denotes the $n$-th order moment about the origin of the pairwise velocity distribution, which is related to the central moments through,
\begin{equation}
m_n = \sum_{k=0}^n {n\choose k} c_k m_1^{n-k}.
\label{eq:central2origin}
\end{equation}
As a result, an approximation to the streaming model is given by,
\begin{equation}
    \xi^S (s_\perp, s_\parallel) \approx \xi^R (s) +
    \sum_n \frac{(-1)^n}{n!} \frac{d^n}{d s_\parallel^n} \left((1 + \xi^R(s)) m_n(\mathbf{s}) \right),
    \label{eq:taylor_expansion}
\end{equation}
where the integral of Eq.~\eqref{eq:streaming} has now been replaced by derivatives of the pairwise velocity moments evaluated at the redshift space position, $\mathbf{s}$.
\Baojiu{Consequently, for large pair separations, where the above approximation works well, 
the exact} shape of the pairwise velocity distribution does not affect the clustering, and it is only the moments of the distribution that influence the redshift space correlation function. \Baojiu{This explains why the Gaussian moments model works so well in Fig.~\ref{fig:multipoles} while the Gaussian ML model, which describes the integrand better, fails to reproduce the multipoles.}

We can use Eq.~\eqref{eq:taylor_expansion} to obtain analytical predictions for the redshift space clustering based on the moments. Up to first order terms the resulting expression is simply
\begin{equation}
\begin{split}
    \xi^{(1)}(s,\mu) \approx &~\xi^R(s) -\frac{d \xi^R(s)}{ds}m_{10}(s)\mu^2 \\
    &- \left(1+\xi^R(s)\right)\left(\frac{m_{10}(s)}{s}(1-\mu^2)+\frac{dm_{10}(s)}{ds}\mu^2\right),
\end{split}
\end{equation}
where $m_{10}=m_{10}(s)$, defined by Eq.~\eqref{eq:moment_def}, denotes the radial mean infall. For the multipoles we have

\begin{equation}
    \xi^{(1)}_0(s) \approx \xi^R -\frac{1}{3}\frac{d \xi^R}{ds}m_{10} - \frac{1}{3}\left(1 + \xi^R\right)\left(2\frac{m_{10}}{s} + \frac{dm_{10}}{ds}\right),
\end{equation}
\begin{equation}
    \xi^{(1)}_2(s) \approx  -\frac{2}{3}\frac{d \xi^R}{ds} m_{10} + \frac{2}{3}\left(1 + \xi^R\right)\left(\frac{m_{10}}{s} - \frac{dm_{10}}{ds}\right),
\end{equation}
\begin{equation}
    \xi^{(1)}_4(s) \approx 0.
\end{equation}
For second order terms, we find 
\begin{equation}
\begin{split}
    \xi^{(2)}(s, \mu) & \approx  \frac{1}{2} \left(\frac{d^2\xi(s)}{d s_\parallel^2} m_2(s_\parallel, s_\perp)  + \left(1 + \xi(s)\right)\frac{d^2m_2(s_\parallel, s_\perp)}{d s_\parallel^2}  \right)  \\
    &+ \frac{d \xi(s)}{d s_\parallel} \frac{d m_2(s_\parallel, s_\perp)}{d s_\parallel}.
    \end{split}
\end{equation}

Note that the analytical result for the second-order multipoles includes many more terms than its first-order equivalent. Therefore, instead of calculating the resulting multipoles analytically, we take numerical derivatives of moments higher than one in the remainder of this work. 

In what follows we address the question of how accurate the expansion Eq.~\eqref{eq:taylor_expansion} is on small scales, and how the different moments affect the clustering multipoles. \Carlton{The expansion} turns out to give accurate predictions for the multipoles even on scales of about $10\,\Mpch$, with the advantage of 
\Baojiu{replacing} the integral in Eq.~\eqref{eq:streaming}, which sums up the contributions of the
\Baojiu{pairwise velocity distributions} on different scales, with a derivative at the scale \Baojiu{$s$} under consideration. Therefore, it converts the non-local relationships between the redshift and real space correlation functions with the pairwise velocity PDF, into a local relation between the redshift space correlation and the derivatives of the pairwise velocity moments and real space correlation function.

In the next subsection, we will test the accuracy of the expansion by comparing it to a full streaming model in which we assume the pairwise velocity distribution is either Gaussian or ST.

\subsection{The range of validity of the streaming model expansion}

Assessing the exactitude of the Taylor expansion on different scales is not straightforward, since including higher order terms involves higher order derivatives of the velocity moments and the real space correlation function.

\Baojiu{As one can see from Fig.~\ref{fig:moments}, on small scales the velocity moments as functions of pair separation measured from the simulation are not smooth and their high-order derivatives can be noisy, which will affect the accuracy of the analytical predictions. Given that our main objective in this subsection is to test the validity of the expansion method, to eliminate the impact of such noise}, we fit the ingredients of the Taylor expansion of the streaming model. The fits to the moments are shown \Baojiu{as dotted lines} in Fig.~\ref{fig:moments}. For the real space correlation function, we fit with a simple power law. \Baojiu{We then take the fitted curves as the ``truth", and} compare the 
\Baojiu{predictions} of the Taylor expansion with two different streaming models in which we can convert the fitted moments into pairwise velocity distributions.
These include a Gaussian streaming model, to demonstrate why the GSM gives accurate predictions on quasi-linear scales where the pairwise velocity distribution is highly non-Gaussian, and an ST model, to show the effect of skewness and kurtosis in improving the accuracy of the expansion. The Gaussian distribution has the correct two lowest central moments, whilst the ST matches the four lowest moments.

We shall not compare the expansion to the simulation measurements directly, since the analytical fitting formulae to the measured moments already induce \Baojiu{at least per cent-level} modifications to the multipoles on small scales. However, this exercise is realistic enough (both the real space correlation function and the moments have been fitted to the N-body simulation results) to demonstrate up to which scale the Taylor expansion method can be used.

\begin{figure*}
\centering
\begin{minipage}[t]{.45\textwidth}
\centering
    \includegraphics[width = \textwidth]{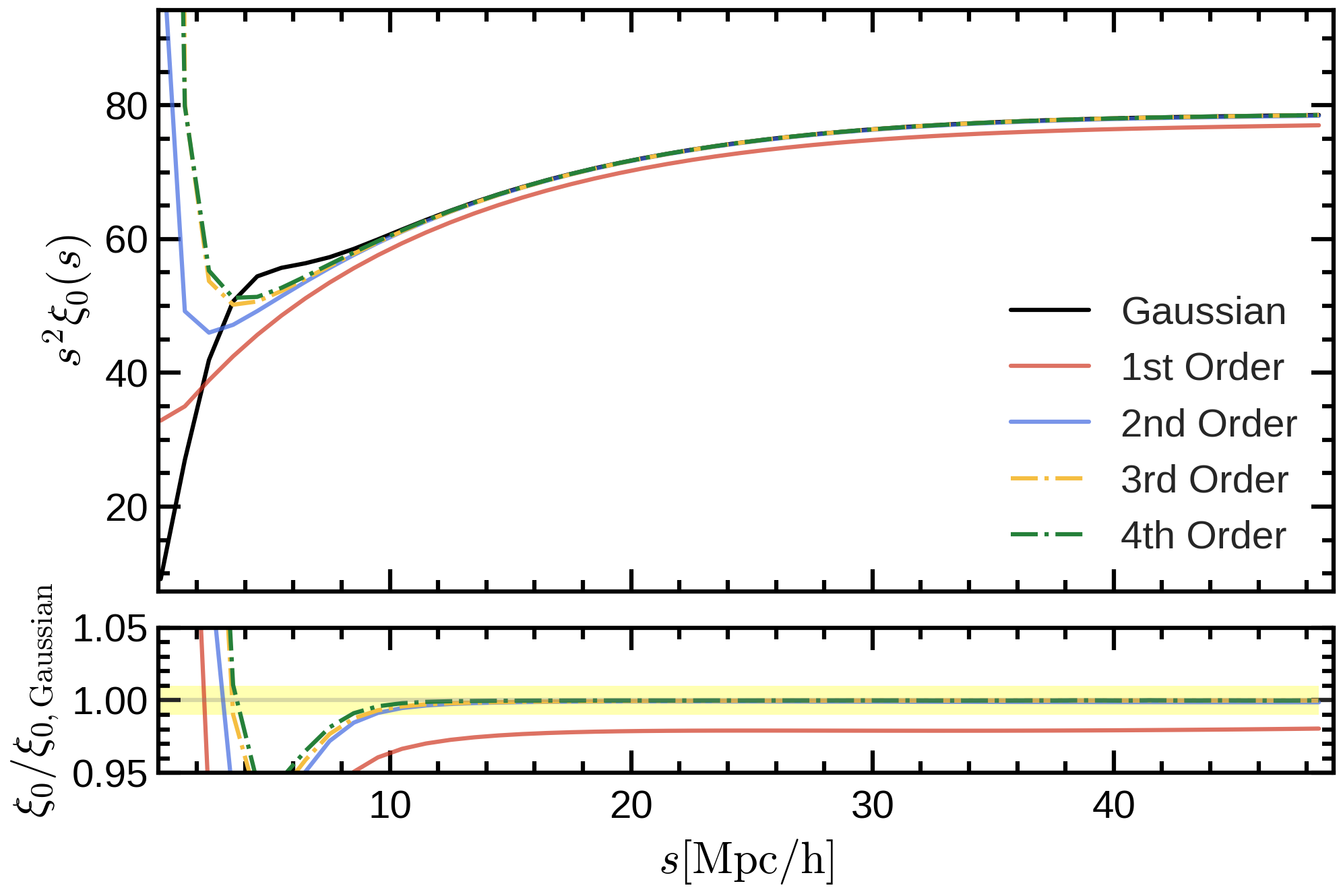}

    \includegraphics[width = \textwidth]{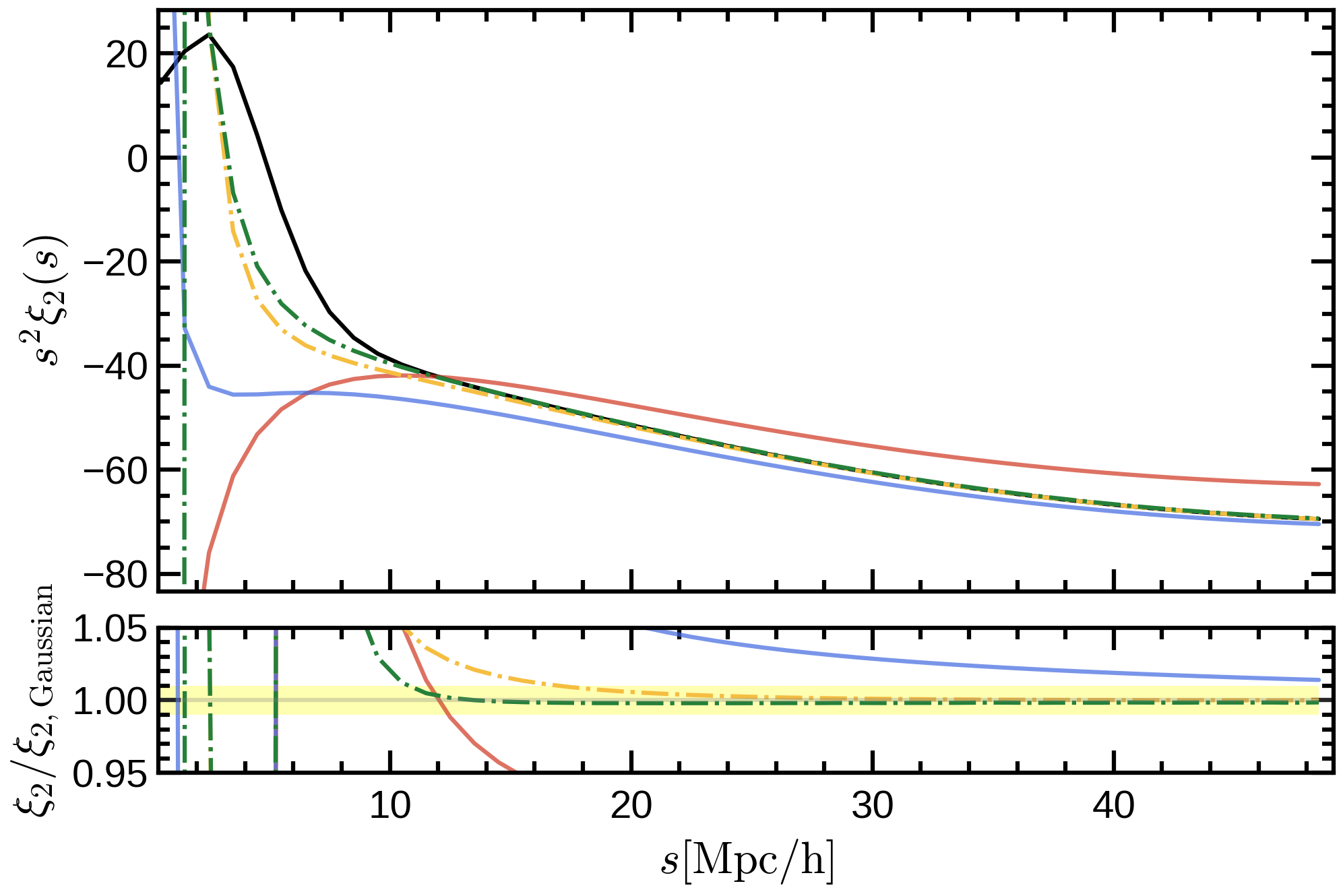}

    \includegraphics[width = \textwidth]{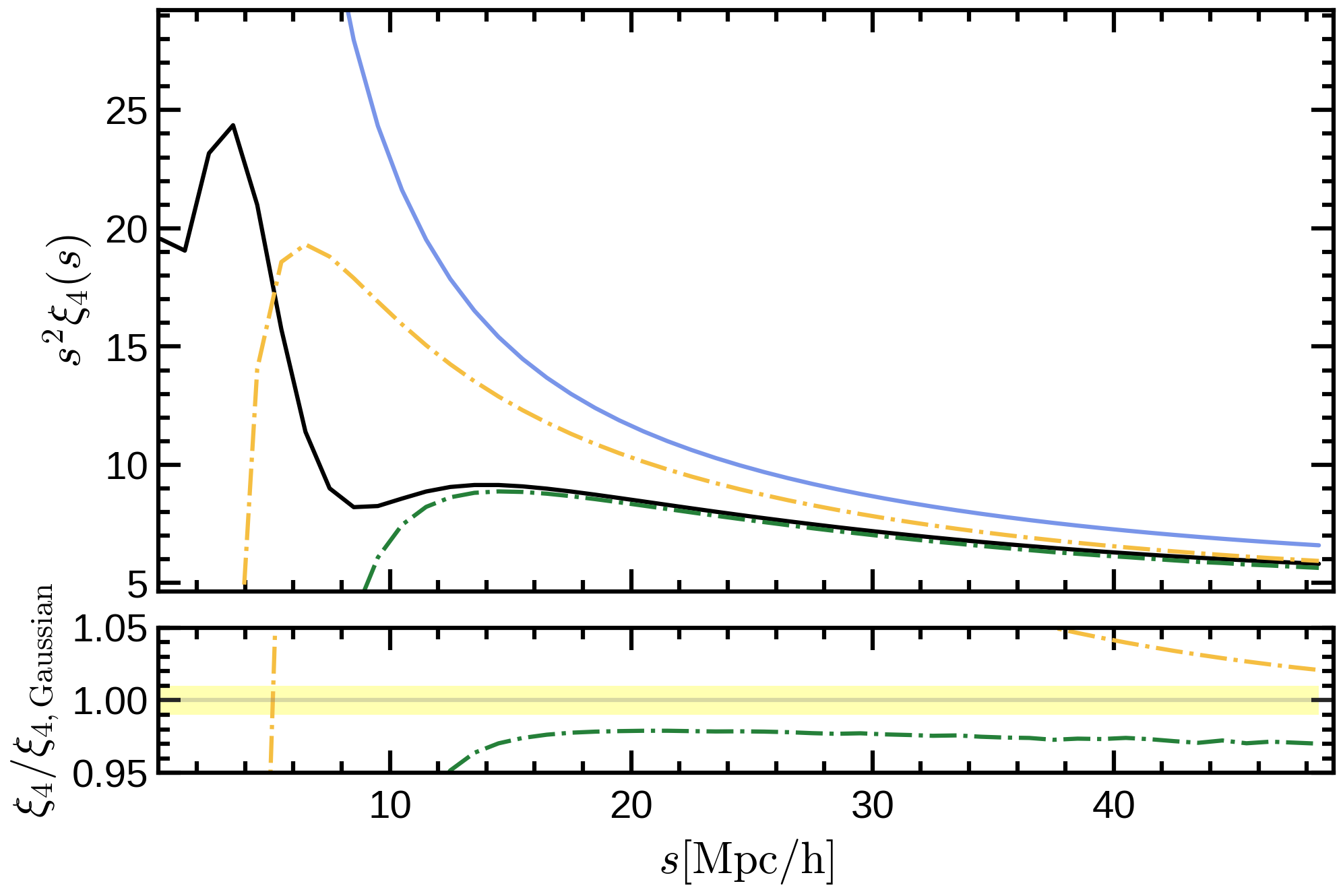}
\caption{Accuracy of the Taylor expansion up to $4$-th order, assuming the pairwise velocity distribution is Gaussian, compared to the full streaming model under the same assumptions. Note all the real space ingredients to the streaming model, the real space correlation function and the two lowest order pairwise velocity moments, are analytical functions fitted to the simulation measurements. The yellow shaded region shows one per cent level agreement between the Taylor expansion and the full Streaming Model. The monopole achieves an accuracy 
\Carlton{better} than the one per cent on scales above $10 \,\, \Mpch$ when the expansion is 
\Carlton{truncated at} second order. For the quadrupole to achieve a similar accuracy, we need to 
\Carlton{retain} up to fourth order terms.}\label{fig:gaussian_toymodel}
\end{minipage}\qquad
\begin{minipage}[t]{.45\textwidth}
    \includegraphics[width = \textwidth]{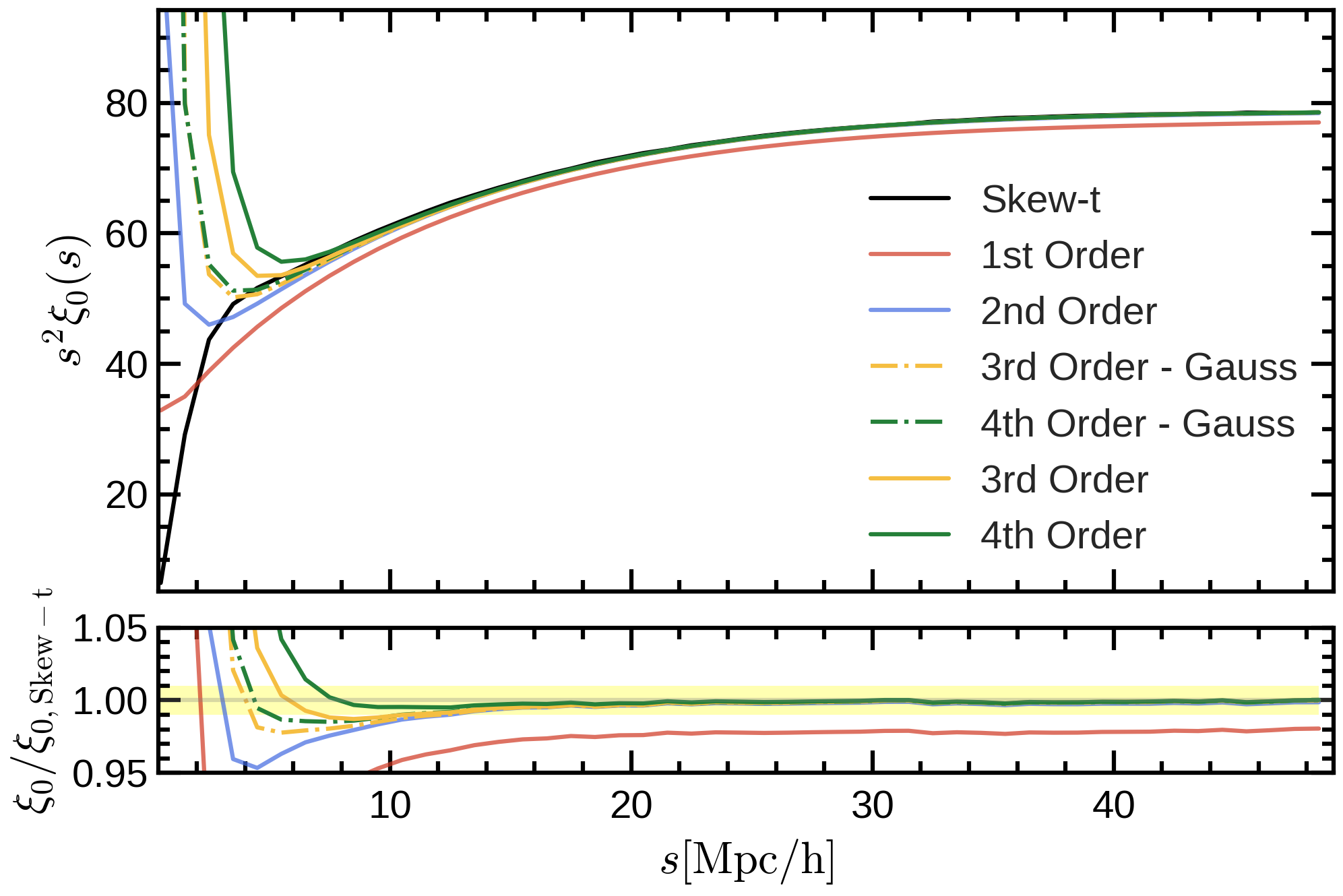}

    \includegraphics[width = \textwidth]{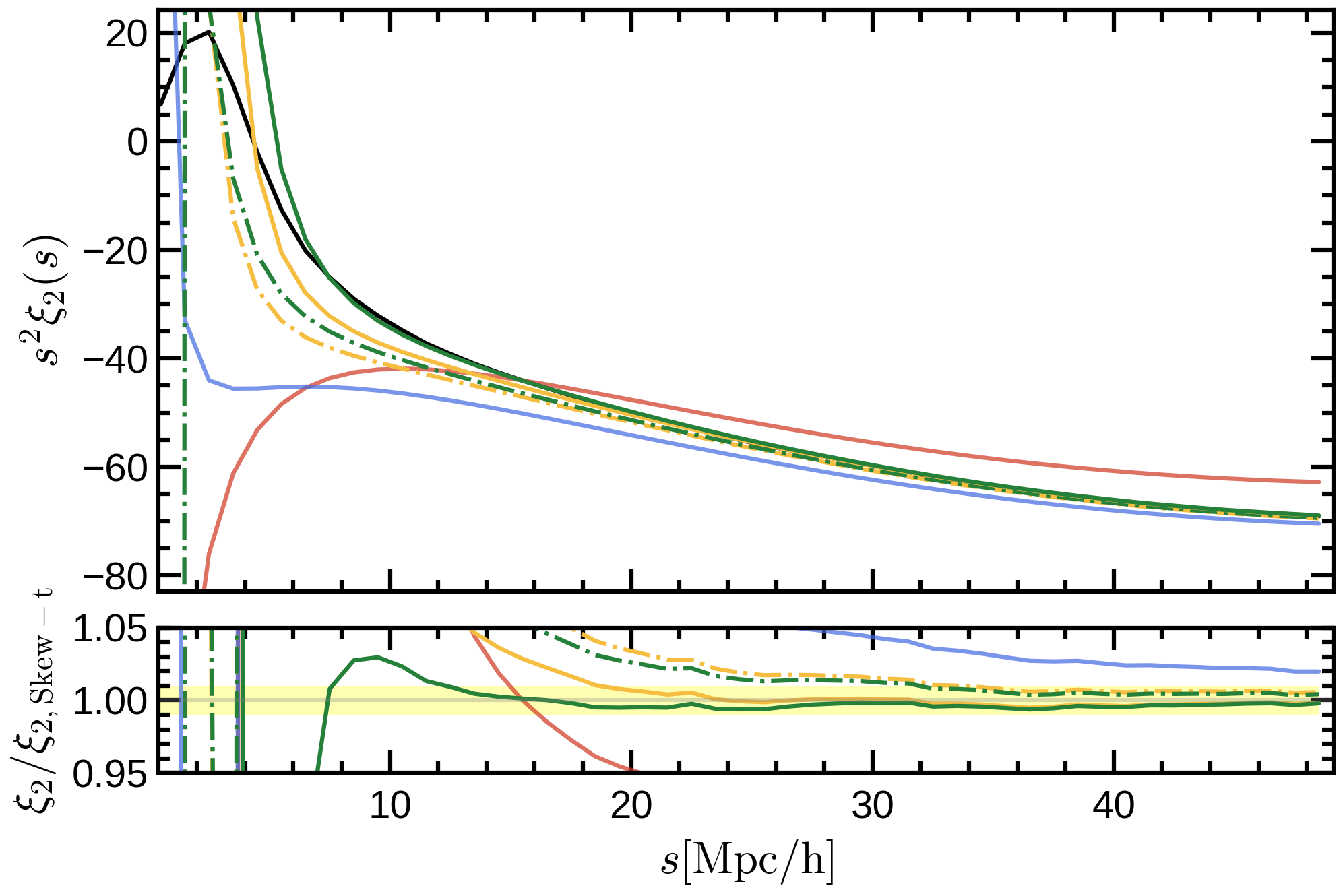}

    \includegraphics[width = \textwidth]{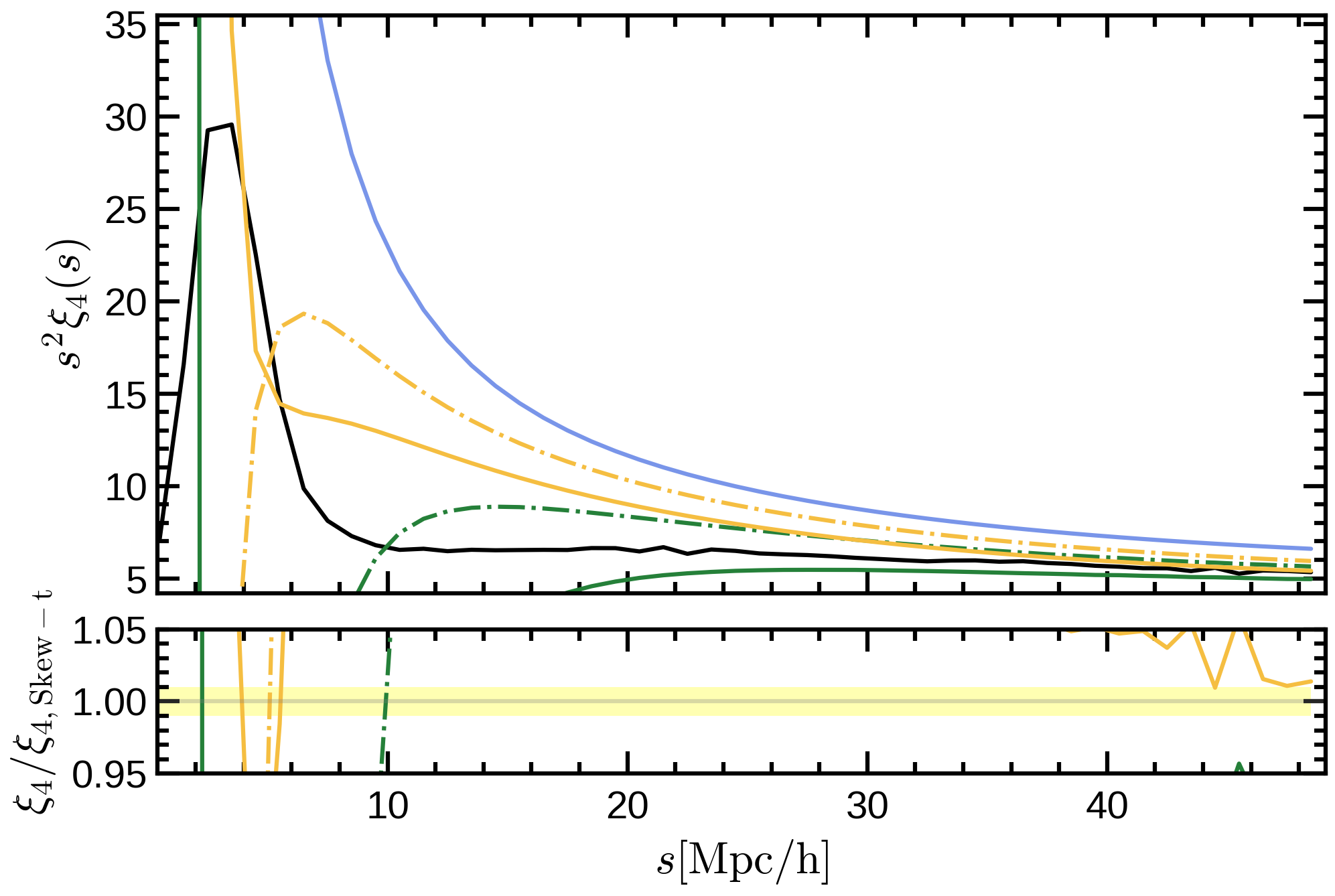}
  \caption{Same as in Fig.~\ref{fig:gaussian_toymodel}, but assuming the pairwise velocity distribution follows a Skew-t distribution. 
  \Carlton{Compared with a} Gaussian distribution, which has the correct first and second order moments, the Skew-t also matches the skewness and kurtosis. For comparison, we show the Taylor expansion found assuming the distribution is Gaussian and the fully non-Gaussian result, in which we include skewness and kurtosis. The effect of the skewness can therefore be seen as the difference between the orange dashed and orange solid lines, whilst the effect of the kurtosis is given by the difference between the green dashed and green solid lines. We find that the effects of the skewness and kurtosis are particularly important for the quadrupole on small scales.}\label{fig:st_toymodel}
\end{minipage}
\end{figure*}

Compared with the full streaming model, the Taylor expansion makes minimal assumptions regarding the pairwise velocity distribution, since it only uses the moments, and removes the integration over all pair separations in Eq.~\eqref{eq:streaming}. For the Gaussian case, the results of expanding Eq.~\eqref{eq:taylor_expansion} up to $n = 4$ are shown \Baojiu{as coloured lines} in Fig.~\ref{fig:gaussian_toymodel}. Note that although for a Gaussian distribution the odd central moments vanish, the odd moments about the origin get contributions from lower order even central moments, as expressed in Eq.~\eqref{eq:central2origin}. Therefore, even orders of the expansion do contribute. \Baojiu{Similarly, while \correction{odd} central moments higher than the second order vanish for a Gaussian distribution, $m_n$ is nonzero for $n>2$. The full streaming model predictions using the integral in Eq.~\eqref{eq:streaming} and a Gaussian pairwise velocity distribution are shown as black sold lines.}

In Fig.~\ref{fig:gaussian_toymodel} we see how including only terms up to $n=2$ we can reproduce the monopole to within $1\%$ 
\Carlton{down} to $10\Mpch$. To achieve a comparable accuracy for the quadrupole, however, we need to add higher order moments up to $n=4$. For the hexadecapole, we expect the Taylor expansion to be less accurate, \Baojiu{because it is more strongly affected by the finger-of-God effect, which originates from the very small and nonlinear scales on which the Taylor expansion breaks down. This is confirmed by the fact that in the lower panel of Fig.~\ref{fig:gaussian_toymodel} there are larger differences between the coloured and black lines.} \Baojiu{Nevertheless, we find that} for the Gaussian model, the hexadecapole predicted by the expansion up to fourth order is accurate to within 
\Baojiu{$3\%$} down to $15\Mpch$.

Finally, since the ST distribution reproduces the measured line-of-sight \Baojiu{velocity distribution} with a higher accuracy than a Gaussian distribution \Baojiu{(Fig.~\ref{fig:line_of_sight_pdf})}, we also demonstrate the effect of higher order moments on the Taylor expansion using an ST model for $\mathcal{P}$. In Fig.~\ref{fig:st_toymodel}  we show both the third and fourth order moment expansion assuming a Gaussian distribution, with zero skewness and fixed kurtosis shown as dashed-dotted lines, and the fully non-Gaussian moments. Although for the monopole, non-Gaussianity does not play an important role, adding the skewness extends the $1\%$ agreement in the quadrupole from scales of around $30\Mpch$ down to $20\Mpch$. The effect of the fourth order moment, kurtosis, is important to extend \Carlton{close agreement} even further to about $10\Mpch$. These results are consistent with the findings in Fig.~\ref{fig:multipoles}, where we find that the ST model improves the agreement of the quadrupole in the range $10$-$30\Mpch$.

\Baojiu{Note that for the hexadecapole (shown in the bottom panel of Fig.~\ref{fig:st_toymodel}) the Taylor expansion method introduces a substantial fractional error of $>5\%$ on all scales, even if the fourth-order corrections are included. This is not surprising because now the assumed true model -- in which the pairwise velocity satisfies an ST distribution -- is more complicated, and because the absolute value of the hexadecapole is much closer to zero which tends to magnify the relative error. Nevertheless, we still observe that including higher-order terms brings the expansion prediction closer to the 
\Carlton{correct answer}. In Fig.~\ref{fig:multipoles}, the multipoles have been numerically calculated using the full Streaming model of Eq.~(\ref{eq:streaming}), rather than the Taylor expansion, under the assumption of the pairwise velocity PDF being either Gaussian or ST.}

\section{Sensitivity to the real space streaming model ingredients}
\label{sec:sensitivity}

In this section, we study the effects of varying the various ingredients -- the real space quantities -- needed to predict the redshift space clustering through the ST streaming model. \Baojiu{This will indicate to us what precision is required for each ingredient in order to make the final prediction of the multipoles accurate.}

\begin{figure*}
\centering
    \includegraphics[width=0.85\textwidth]{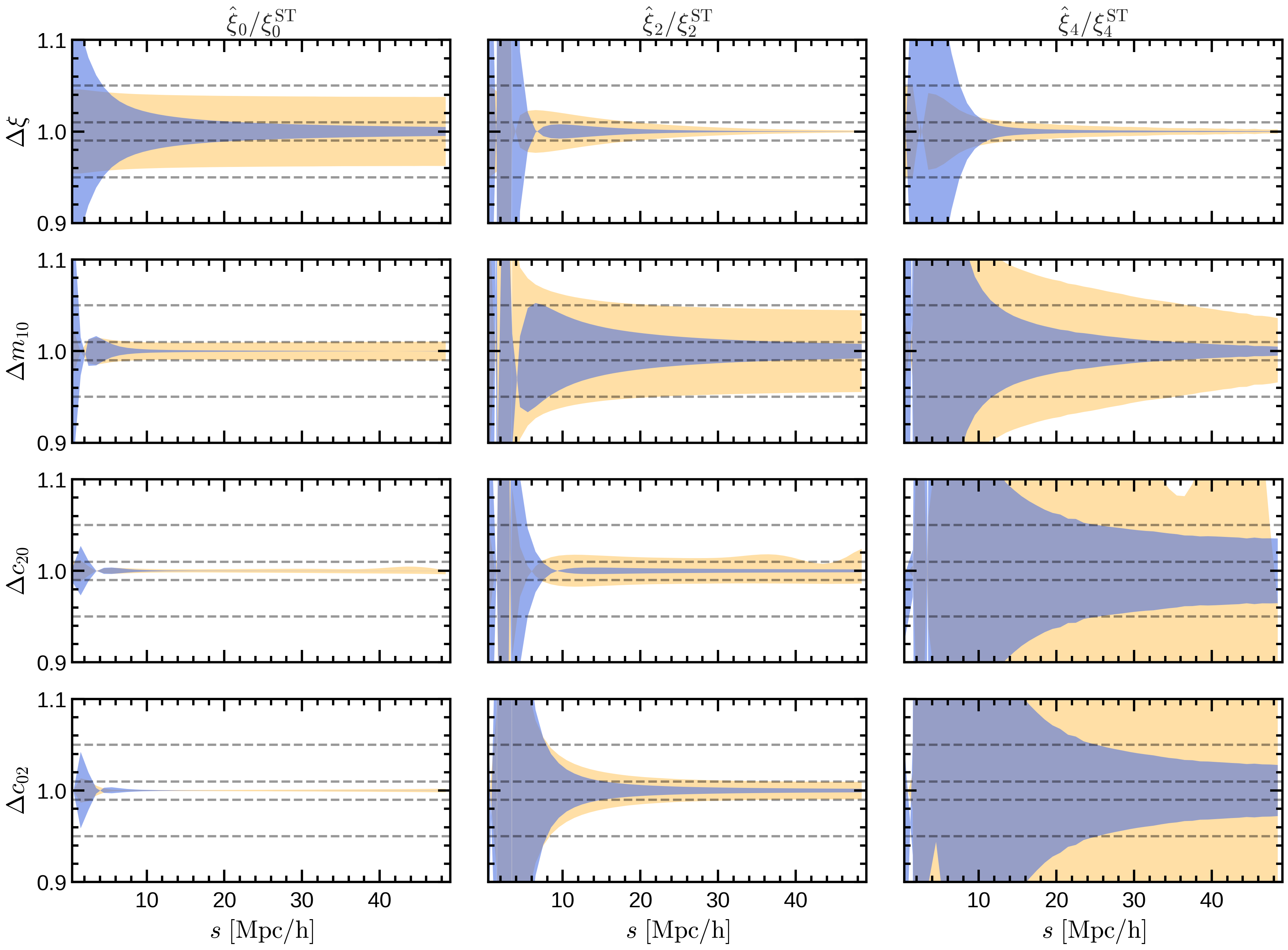}
    \caption{The fractional variation in the monopole, quadrupole and hexadecapole after modifying the real space ingredients of the streaming model. In each row we show the effect of varying: the real space correlation function, the mean pairwise velocity, the radial variance of the pairwise velocity and the transverse variance of the pairwise velocity. Orange contours show the effect of varying 
    \Carlton{each of these quantities} by $\pm 5\%$, while the blue contours vary them by a percentage that depends on scale and increases with $1/r$. On small scales, where perturbation theory predictions degrade, we vary each of the ingredients by a larger percentage. The variation is tuned to produce a $5 \%$ change on scales of $5\Mpch$ and a $1\%$ one on scales of $30\Mpch$. Gray dashed lines determine $5$ and $1$ per cent deviations from the true model.}
    \label{fig:varying_low_order}
 \end{figure*}

In Fig.~\ref{fig:varying_low_order}, we show the effects on the multipoles of varying the real space correlation function \Baojiu{(the first row)}, the mean pairwise velocity \Baojiu{(the second row)} and its variance (both in the radial and transverse components; \Baojiu{the last two rows}). We have studied the impact of two types of variations: a constant change by $\pm 5\%$, or a fractional change that increases towards smaller pair distances as $1/r$, to emulate the fact that perturbation theory predictions worsen towards small scales. The latter gradual change is tuned to vary the given function by \Baojiu{$\pm5\%$} on scales of $5\Mpch$ and by \Baojiu{$\pm1\%$} on the scale of $30\Mpch$. In this way we can compare the effect of a varying slope due to uncertainties in our predictions, which we know is important since derivatives of the moments appear in the Taylor expansion Eq.~\eqref{eq:taylor_expansion}. \Baojiu{The fractional changes to the predicted redshift-space monopole, quadrupole and hexadecapole (from left to right) are respectively shown as orange and blue shaded regions for the constant and scale-dependent changes.}

Varying the real space correlation function by $5\%$ produces approximately the same 
\Baojiu{fractional} change in the monopole. Since the $0$-th order contribution to the quadrupole and hexadecapole is zero, \Baojiu{both the constant and the scale-dependent variations in} the real space correlation function produce a sub-per cent effect on the quadrupole and hexadecapole on scales larger than $20\Mpch$.

The mean pairwise velocity has a stronger effect on both the quadrupole and hexadecapole. The importance of getting the slope of the mean right, found in the Taylor expansion, can also be seen in the blue contours. A change of $1\%$ above $30\Mpch$ produces a slightly larger effect on the quadrupole. \Baojiu{The monopole, in contrast, is much less sensitive to variations of the mean pairwise velocity, and the effect is 
\Carlton{ at the sub per cent level} at $s\gtrsim10\Mpch$ for both variation scenarios.}

On the other hand, the hexadecapole is most sensitive to the radial and transverse standard deviations. Changes of $5\%$ 
\Carlton{can produce a change that is twice as large in the hexadecapole}. 

Regarding the third-order moments, we 
found in the last section (Fig.~\ref{fig:st_toymodel}) that the skewness 
\Baojiu{has at most} a per cent\Baojiu{-level effect on the monopole and quadrupole on} 
scales below $30\Mpch$. Since its effect is very small, we do not show the equivalent in Fig.~\ref{fig:varying_low_order}. We find that varying the third order radial and transverse moments by $50\%$ introduces modifications smaller than $5\%$ on the quadrupole on small scales. 

Finally, the effect of fourth order terms is important on scales below $20\Mpch$. However, we have already shown  in the previous section that setting the fourth order moments to zero, by assuming Gaussianity, also gives only a few percentage level corrections to the quadrupole on small scales (see difference between orange dashed line and solid green line in Fig.~\ref{fig:st_toymodel}).

Therefore, even on small scales, we need to predict most accurately the lower order moments: the mean and the standard deviation, particularly the latter, if we want to utilise information contained in the hexadecapole. We can afford to have a larger margin of error on the predictions of the higher order moments, 
and still extend the validity up to scales of around $10\Mpch$.

\section{Conclusions and Discussion}

The new generation of surveys \citep{Amendola2013,2014PASJ...66R...1T,Levi:2019ggs,10.1117/12.2312012} is going to measure redshift space clustering of galaxies with unprecedented precision. To translate the high accuracy of these measurements into tighter constraints on the cosmological parameters or on possible deviations from general relativity, we need to improve our theoretical models of redshift space distortions (RSD). Within the streaming model of RSD, we need to: i) improve the mapping from real to redshift space, i.e., by developing the modelling of the pairwise velocity \Baojiu{distribution} including its higher order moments, ii) increase the accuracy of the predictions of the 
ingredients of the streaming model -- the real space correlation function and the pairwise velocity moments \Baojiu{-- for given cosmological parameters}. 
\Carlton{Here}, we have focused on the first \Carlton{ of these} aspects, but we have also briefly analysed the effects of the second.

In N-body simulations, where the fully non-linear evolution of 
\Carlton{collisionless} particles is solved, we observe that the \Baojiu{distribution of the} pairwise velocities of dark matter halos is skewed towards negative velocities, and has 
\Carlton{broader} tails than a Gaussian. Therefore, 
models that use Gaussian distributions do not give an accurate description of the pairwise velocities. We have introduced an extension to the Gaussian streaming model by using the Skew-T probability distribution \Baojiu{for the pairwise velocity}. 
\Carlton{The parameters of this distribution} 
can be tuned to match the four lowest-order velocity moments \Carlton{measured} from simulations. The ST model describes the simulation measurement of the pairwise velocity distribution significantly better than a simple Gaussian.

We compare two different methods to find the 
\Carlton{best-fitting} parameters of the 
\Baojiu{pairwise velocity} distribution: maximum likelihood estimation and the method of moments. Although \Baojiu{the results of} both 
\Carlton{approaches} seem to describe the 
\Baojiu{measured velocity} distribution equally well on large scales, they give very different results for the redshift space clustering once inserted into the streaming model. Using the method of moments is crucial for describing all multipoles, including the small scales. Even though the Gaussian distribution gives a very poor fit to the measured pairwise velocities distribution, it can reproduce the 
true multipoles on quasi-linear scales within the \Baojiu{small} statistical errors of our simulations when we tune it to 
have the 
\Baojiu{two lowest-order moments extracted from the simulations. On the other hand}, the best-fit Gaussian found by maximising the likelihood gives results that are more than five standard deviations away from the 
\Baojiu{simulation measurement}.
 
The ST model, also using the method of moments, gives predictions for the redshift space multipoles (monopole, quadrupole and hexadecapole) that are within the \Baojiu{small} statistical \Carlton{sampling variance} errors   
\Carlton{(driven by the simulation volume)} 
\Baojiu{down} to about $10\Mpch$. On \Baojiu{such small} scales, the Gaussian \Baojiu{streaming} model gives predictions that are more than five standard deviations away from the mean measurement \Baojiu{from simulations}. Therefore, the ST model extends the validity of the streaming model from $30\Mpch$ to $10\Mpch$, and gives a more accurate description of the hexadecapole, which \Baojiu{has so far not been} used in analyses that rely on the Gaussian streaming model \citep[e.g.,][]{2017MNRAS.469.1369S,10.1093/mnras/sty506}, due to its 
\Baojiu{poor} accuracy.

We have used a Taylor expansion of the integrand to show why the Gaussian streaming model can reproduce the clustering on quasi-linear scales within the error bars of the simulation measurement, despite giving a poor description of the pairwise velocity distribution. \Baojiu{At $s\gtrsim30\Mpch$}, only the first and second order moments, the mean and the standard deviation, of the pairwise velocity distribution, \Baojiu{are crucial for determining} the monopole and quadrupole of the two-point correlation function \Baojiu{in redshift space}.

We have also shown that the Taylor expansion can describe the non-Gaussian ST streaming model \Baojiu{down} to smaller scales, of about $10\Mpch$, when expanded up to fourth order. The main advantage of the Taylor expansion is that it 
\Baojiu{replaces the} integral of the pairwise velocities over all scales by a derivative of the moments at the scale under consideration. It therefore makes no assumptions about the \Baojiu{details of the} underlying 
\Baojiu{velocity distribution}, and can give analytical predictions for the monopole and quadrupole. However, it cannot reproduce the hexadecapole as accurately as the \Baojiu{full ST streaming model integral, Eq.~\eqref{eq:streaming}, nor is it as accurate on smaller scales, $s\lesssim15\Mpch$}.

The Taylor expansion could be 
\Carlton{particularly} useful to measure the velocity moments from the observed redshift space multipoles, as was already proposed by \cite{2015MNRAS.446...75B}, along the line of previous measurements of the pairwise velocity dispersion
\citep{2006MNRAS.368...37L, 2018MNRAS.474.3435L}. The main difficulty to measure the pairwise distribution from observations lies in the pair distance dependence of the moments, imprinted by gravity. We would need to develop analytical formulae to summarise the pair distance dependence in a small set of parameters, that are valid independently of the underlying model of gravity. These parameters could then be inferred from observations of redshift space clustering, by running a Monte Carlo Markov Chain. The direct measurement of the moments could be a complementary test of gravity to the growth rate, and it would utilise more information of the full scale dependence of different gravity models. 

Finally, we 
\Baojiu{qualitatively} analysed the 
\Baojiu{effects of inaccurate knowledge} of the real space correlation function or of the velocity moments on the \Baojiu{predictions of the} redshift multipoles. As expected, the monopole is mainly determined by the real space correlation function. We have shown that perturbation theory based CLEFT method \citep{2016JCAP...12..007V} produces per cent-\Baojiu{level-accuracy} 
predictions of the real space correlation function. However, to obtain per cent-level accurate predictions for both the monopole and quadrupole, we also need 
\Baojiu{similar accuracy} for the mean pairwise velocity and its slope. Fitting the CLEFT predictions, with five free parameters, we were only able to obtain 
\Carlton{predictions accurate at the per cent level} 
for the mean on scales above  $35\Mpch$. On the other hand, the hexadecapole is very sensitive to the variance of pairwise velocities, \Baojiu{for which} CLEFT is only accurate to one per cent above scales of about  $45\Mpch$. Therefore, future efforts to utilise the information content in the hexadecapole will have to obtain more accurate theoretical prescriptions for the variance. Per cent level errors on the prediction of the variance become even larger errors on the hexadecapole. On scales smaller than $30\Mpch$, we also need predictions for the skewness and kurtosis of pairwise velocities. However, these do not need to be as accurate: per cent errors on the skewness and kurtosis have negligible impact on the multipoles.

To summarise, we have developed a streaming model based on the Skew-t distribution \Baojiu{of pairwise velocities}, that accurately describes redshift space clustering on scales larger than $10$-$15\Mpch$, given the first four 
moments of the pairwise velocity distribution are known. In order to improve constraints on the growth rate by using the ST model, we need to improve the theoretical predictions of the pairwise velocity moments and its dependency on the cosmological parameters. In this work, we have focused our analysis on massive dark matter haloes at redshift zero, and we leave the study of galaxies at a range of different redshifts 
to future work.

\section*{Data availability}
 The data shown in this article will be shared on reasonable request to the corresponding author.
\section*{Acknowledgements}
We would like to thank Davide~Bianchi, Ravi~Sheth, Emanuele~Castorina, Joseph~Kuruvilla and Cristiano~Porciani for useful discussions. 
CCL is supported by a PhD Studentship from the Durham Centre for Doctoral Training in Data Intensive Science, funded by the UK Science and Technology Facilities Council (STFC, ST/P006744/1) and Durham University. BL, PZ and CMB acknowledge support from the STFC through ST/P000541/1. BL and AE acknowledge the support of the European Commission through the European Research Council (ERC-StG-PUNCA-716532).
This work used the DiRAC@Durham facility managed by the Institute for Computational Cosmology on behalf of the STFC DiRAC HPC Facility (www.dirac.ac.uk). The equipment was funded by BEIS capital funding via STFC capital grants ST/K00042X/1, ST/P002293/1 and ST/R002371/1, Durham University and STFC operations grant ST/R000832/1. DiRAC is part of the National e-Infrastructure. 
This work was supported in part by World Premier International Research Center Initiative (WPI Initiative), MEXT, Japan, JSPS KAKENHI Grant Numbers JP15H03654, JP15H05887, JP15H05893, JP15H05896, JP15K21733, JP17K14273, and JP19H00677, and Japan Science and Technology Agency CREST JPMHCR1414. TN and MT acknowledge the supported by the Munich Institute for Astro- and Particle Physics (MIAPP) of the DFG cluster of excellence ORIGINS (http://www.munich-iapp.de).
Numerical simulations presented in this paper were carried out on Cray XC30 and XC50 at Centre for Computational Astrophysics, National Astronomical Observatory of Japan.

\bibliographystyle{mnras}
\bibliography{references}

\begin{appendices}
\section{Method of moments for the ST distribution}
\label{app:moments_st}
The four parameters of the ST distribution ($v_c, w, \alpha, \nu$) are determined by the first four order moments. To simplify the relation between moments and parameters, we introduce,
\begin{equation}
b_\nu = \left( \frac{\nu}{\pi} \right)^\frac{1}{2} \frac{\Gamma \left( \frac{\nu - 1}{2} \right)}{\Gamma(\nu/2)},
\end{equation}\begin{equation}
\delta = \frac{\alpha}{\sqrt{(1 + \alpha^2)}},
\end{equation}The moments are therefore given by,
\begin{equation}
m_1 = v_c + w \delta b_\nu,
\end{equation}
\begin{equation}
c_2 = w^2 \left( \frac{\nu}{\nu - 2} - \delta^2 b_\nu^2 \right),
\end{equation}
\begin{equation}
\gamma_1 = \frac{c_3}{c_2^{3/2}} = \delta b_\nu 
\left( \frac{\nu ( 3 - \delta^2)}{\nu - 3} - \frac{3 \nu}{\nu -2} + 2 \delta^2 b_\nu^2  \right) \left(\frac{\nu}{\nu-2} - \delta^2 b_\nu^2\right)^{-\frac{3}{2}},
\end{equation}\begin{equation}
\begin{split}
\gamma_2 = &\frac{c_4}{c_2^2} = \left( \frac{3 \nu^2}{(\nu -2)(\nu-4)}     
					- \frac{4 \delta^2 b_\nu^2 \nu (3 - \delta^2)}{\nu - 3}
                    - \frac{6 \delta^2 b_\nu^2 \nu}{\nu - 2} - 3 \delta^4 b_\nu^4 \right)\\
                    & \left(\frac{\nu}{\nu -2} - \delta^2 b_\nu^2 \right)^{-2} .
 \end{split}
\end{equation}
The parameters $\alpha$ and $\nu$ are obtained from the last two equations that determine the skewness and kurtosis of the distribution, these form a system of non-linearly coupled equations that we solve numerically. The remaining two parameters, $v_c$ and $w$, can then directly be obtained from the equation for the mean and the variance.

\section{Zoom in distributions}
\label{app:zoom_in}

\begin{figure*}
    \centering
    \includegraphics[width=0.9\textwidth]{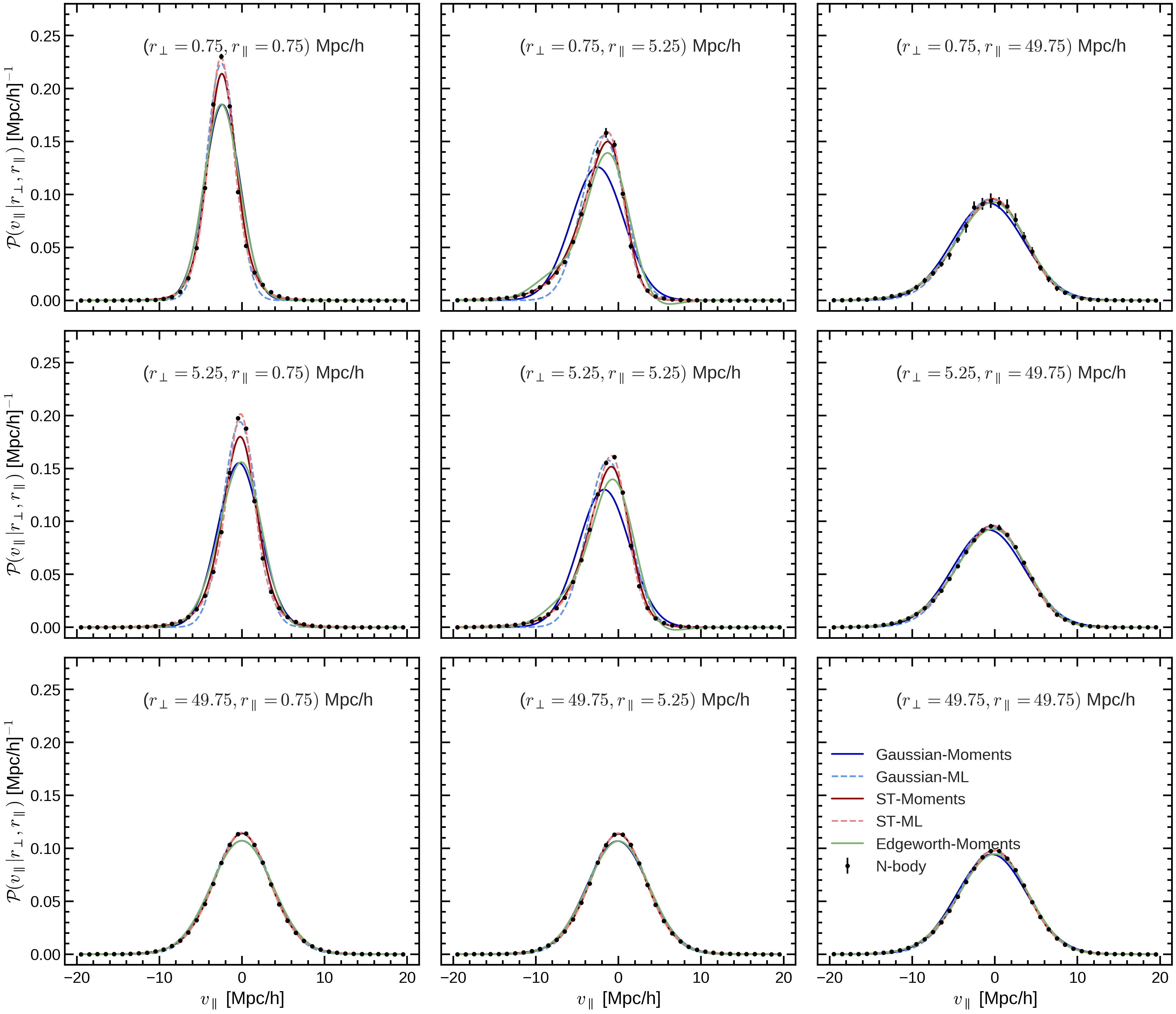}
    \caption{\correction{Linear scale representation of the pairwise velocity distribution to highlight the behaviour of the PDF close to its peak. The models shown are the same one as in Fig.~\ref{fig:line_of_sight_pdf}. Note that the Edgeworth expansion predicts negavite probabilities for certain 
    pair separations such as
    $(r_\perp = 0.75, r_\parallel=5.25) \Mpch$ and $(r_\perp = 5.25, r_\parallel=5.25) \Mpch$, 
where the skewness is more pronounced. Moreover, 
an Edgeworth expansion behaves very differently from a Taylor expansion since it produces an asymptotic expansion, and therefore adding more terms does not guarantee convergence. See \citet{Sellentin:2017aii}
for an interesting discussion on the Edgeworth expansion and its applications to cosmology.
In the application of the Edgeworth expansion to the pairwise velocity distribution, we see that it does not reproduce the N-body measurements as well 
as the ST distribution does with only one extra parameter.}}
  \label{fig:line_of_sight_pdf_nolog}
\end{figure*}

\section{Perturbation Theory results in detail}
\label{app:pt_details}

In this Appendix, we show a detailed summary of the state-of-the-art CLPT and CLEFT perturbation theory predictions for the Gaussian Streaming Model ingredients. Note that we show the predictions for real space statistics, since we want to separately analyse the accuracy of perturbation theory predicting the ingredients of the Streaming Model, and the assumption of a Gaussian pairwise velocity distribution.

The free parameters are found by maximising the combined Gaussian likelihood that the simulation measurements are most probable under the given theory,
\begin{equation}
\log(\mathcal{L}) = \log(\mathcal{L_\xi}) + \log(\mathcal{L}_{m_{10}}) + \log(\mathcal{L}_{c_{20}})   + \log(\mathcal{L}_{c_{02}})   ,
\end{equation}
where the individual likelihoods are given by,
\begin{equation}
\log(\mathcal{L}_y) = - \frac{1}{2} \sum_i \frac{(y_{i, \mathrm{measured}} - y_{i, \mathrm{model}})^2}{\sigma^2_i}.
\end{equation}
where $y$ is the mean simulation measurement across the 15 independent simulations, and $\sigma$ its standard deviation. Note that the covariance matrix is assumed to be diagonal, which means that the parameter uncertainties obtained from the fit will be underpredicted. \correction{While this assumption will also affect the values of the best-fit parameters in detail, we do not expect this to have a qualitative impact on the relative agreement between the model predictions and data, which is our main objective here.} We maximise the likelihood in the pair separation range $15 \, \Mpch < r < 150 \Mpch$ and the resulting mean parameter values are shown in Table~\ref{table:pt_params}. We find a value for the second order Lagrangian bias $b_2$ that is in good agreement with previous measurements \citep{2016JCAP...02..018L}, whereas the tidal bias is rather different from its local Lagrangian value $(b_s = 0)$, which is in contrast with other analyses in the literature \citep{LazSch0918,AbiBal0618}. We also note that the EFT parameters are the least constrained by our measurements, which is to be expected as they only have an impact on the small-scale regime. 
\begin{table*}
\begin{center}
 \begin{tabular}{c c c c c c c c} 
 \hline
   & $b_1$ & $b_2$ & $b_s$ & $\alpha_\xi$ & $\alpha_v$  & $\sigma_{\mathrm{FoG}}$ \\ [0.5ex] 
 \hline\hline
 CLPT & $0.29 \pm 0.01$ & $-1.63 \pm 0.31$ & $ 1.80 \pm 0.38$ & - & - & $-17.35 \pm 0.15$\\ 
 \hline
  CLEFT & $0.30 \pm 0.02$ & $-1.69 \pm 0.26$ & $ 2.16 \pm 0.37 $ & $-39.58 \pm 16.32$  & $90.30 \pm 73.68$ & $-17.45 \pm 0.28$\\ 
 \hline
\end{tabular}
\end{center}
\caption{Perturbation theory parameters for both CLEFT and CLPT. Note that $b_1$, $b_2$, and $b_s$ are obtained by expanding the bias function in Lagrangian space. We show the maximum likelihood estimate and errors representing 1-sigma deviations in the posterior distribution of the given parameter. }
\label{table:pt_params}
\end{table*}

\begin{figure}
\centering
    \includegraphics[width=0.45\textwidth]{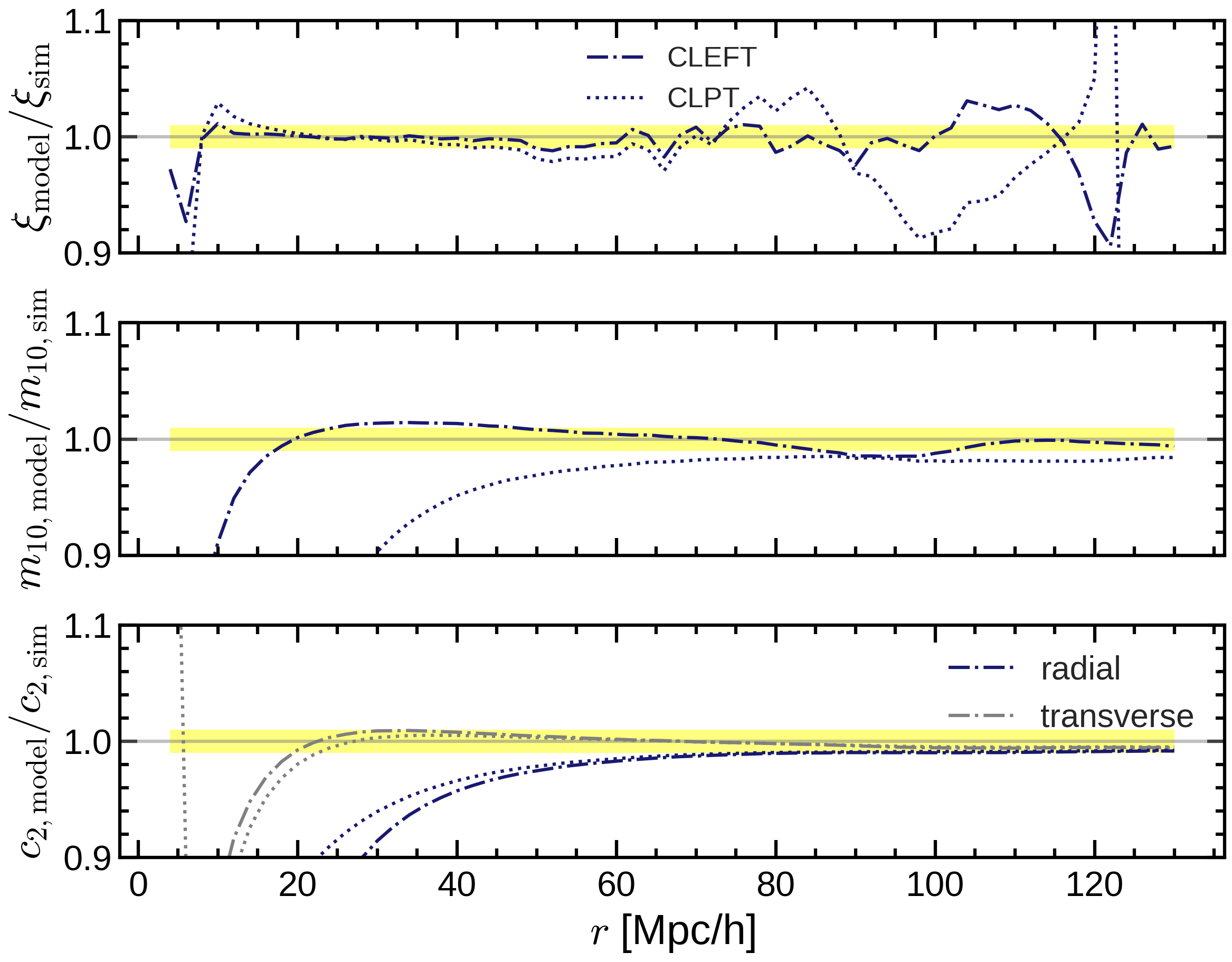}
\caption{Detatiled comparison of the different predictions for the Gaussian Streaming Model ingredients made by CLPT and CLEFT. The top panel shows the ratio of the predicted two-point correlation function to the measurement in the simulation, for both CLPT (dotted) and CLEFT (dotted-dashed line). The solid yellow bands marks the one per cent agreement. The middle and bottom panels show the same comparison for the mean radial velocity, and the second order radial and transverse moments.}
\label{fig:ratios_pt}
 \end{figure}

In Fig~\ref{fig:ratios_pt} we show a detailed comparison of the best-fit model predictions for the two methods. The second counter-term introduced in CLEFT improves notably the prediction for the mean pairwise velocities on scales between $20 \, \Mpch$ and $60 \, \Mpch$. Regarding the second order moments, the predictions for $m_{20}$ are similar for CLPT and CLEFT, however, since $c_{20} = m_{20} - m_{10}^2$, the variance of the radial component is influenced by the predictions of the mean. Conincidentally, the error made by CLPT in the mean improves the agreement with the variance of the radial component (dotted blue line in the lowest panel).

\begin{figure}
\centering
    \includegraphics[width=0.45\textwidth]{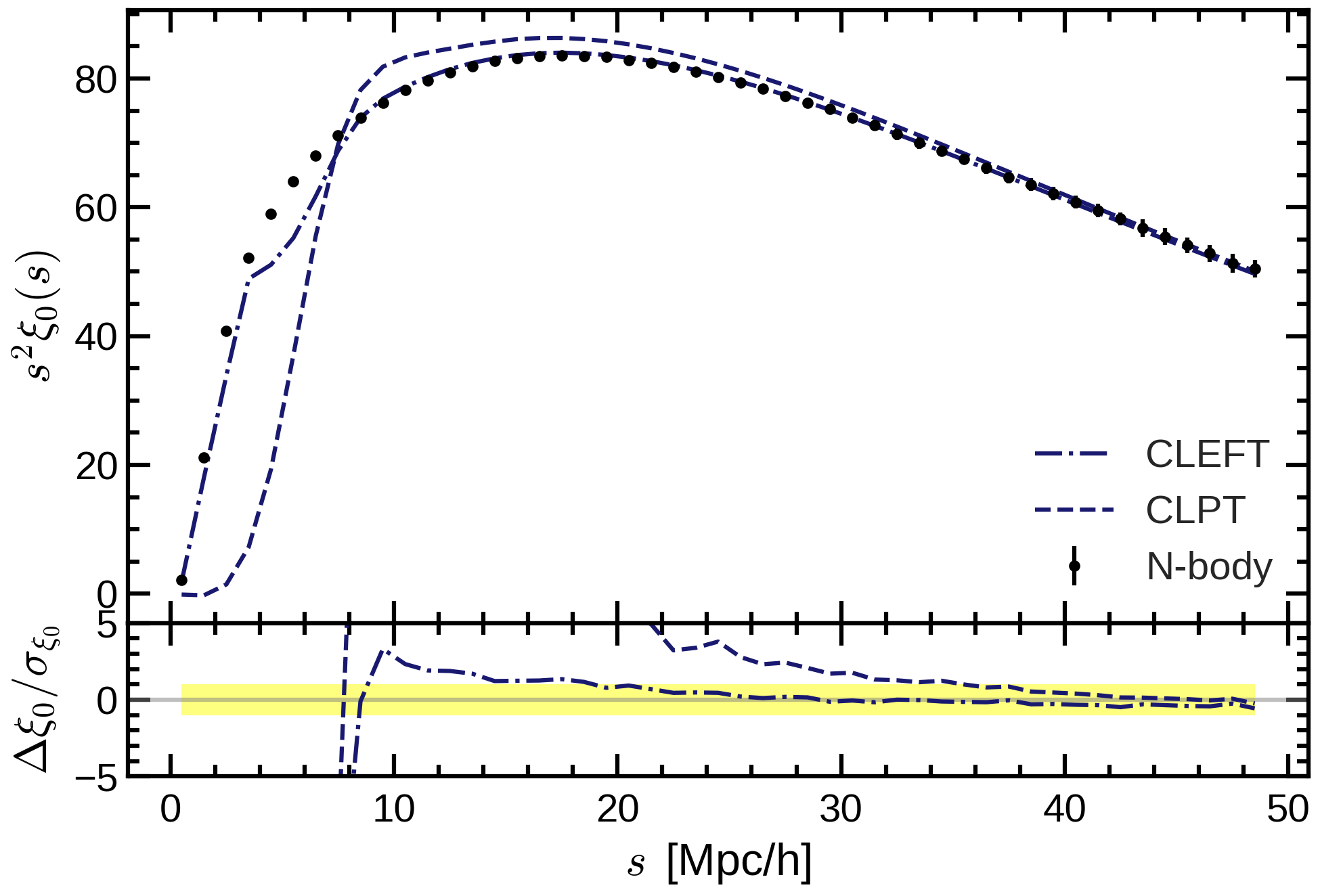}

    \includegraphics[width=0.45\textwidth]{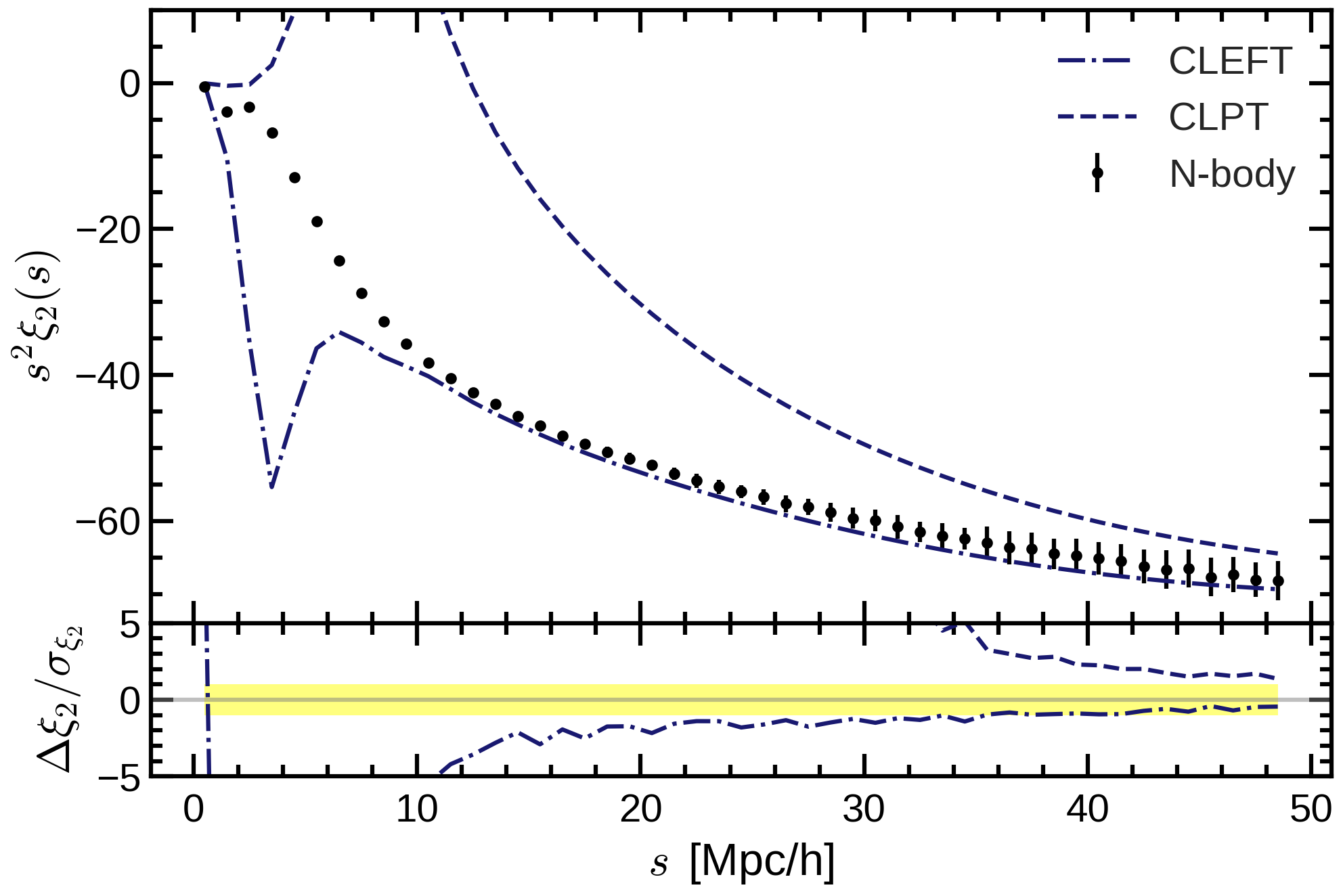}

\caption{Comparison of the Gaussian Streaming model predictions for the redshift space monopole and quadrupole, using the real space ingredients predicted by CLPT and CLEFT. The residuals are plotted as the difference between the model and the simulation in units of the variance calculated across the different independent simulations. The yellow bands show the 1$\sigma$ deviation. }
\label{fig:multipoles_perturbation_theory}
\end{figure}

Finally, we show the redshift space monopole and quadrupole in Fig.~\ref{fig:multipoles_perturbation_theory}, obtained by combining these predictions with the Gaussian Streaming Model. The CLEFT predictions of the monopole and quadrupole are more accurate than those from CLPT, mainly due to the increased accuracy in estimating the mean pairwise velocity, which is consistent with our findings in Sec.~\ref{sec:sensitivity}. As shown in \cref{sec:redshift_space}, on scales smaller than $30 \Mpch$ it is necessary to include higher order moments to further improve the accuracy of the predictions. A more detailed comparison of these different models applied to mock catalogues that mimic actual data at different redshifts and different halo mass ranges will be the subject of future work.

\end{appendices}

\bsp	
\label{lastpage}
\end{document}